\documentclass[usenatbib,twocolumn]{mnras}
\pdfoutput=1

\usepackage{mathptmx}
\usepackage[T1]{fontenc}
\usepackage{ae,aecompl}
\usepackage{times}
\usepackage{amssymb}
\usepackage{amsmath}
\usepackage{upgreek}
\usepackage{color}
\usepackage{pdflscape}
\usepackage{multicol} 
\usepackage{amsmath}
\usepackage{caption}
\usepackage[demo]{graphicx}
\usepackage{subfig,lipsum}
\usepackage{float}
\usepackage{subfig}
\usepackage{siunitx}
\usepackage{fancyhdr}
\usepackage{changepage}
\usepackage{placeins}
\usepackage{ulem}
\usepackage{threeparttable}

\DeclareCaptionLabelFormat{cont}{#1~#2\alph{ContinuedFloat}}
\usepackage{newtxtext,newtxmath}

\usepackage[T1]{fontenc}
\DeclareUnicodeCharacter{0080}{$^$}
\title[Polarization versus extinction and distance]{Revealing the dust grain polarization properties as a function of extinction and distance towards NGC\, 1893}

\author[Bijas et al.]{N. Bijas,$^{1}$\thanks{Email: bijasthejas@gmail.com, bijasn@students.iisertirupati.ac.in}
Chakali Eswaraiah,$^{1}$\thanks{Ramanujan Fellow; Email: eswaraiahc@labs.iisertirupati.ac.n}
Jia-Wei Wang,$^2$ 
Jessy Jose,$^1$ 
Wen-Ping Chen,$^3$
Di Li,$^{4,5}$
Shih-Ping Lai,$^6$ 
\and D. K. Ojha $^7$\\
\\
$^{1}$Indian Institute of Science Education and Research (IISER) Tirupati, Rami Reddy Nagar, Karakambadi Road, Mangalam (P.O.), Tirupati 517 507, India\\
$^{2}$ Academia Sinica Institute of Astronomy and Astrophysics, No. 1, Section 4, Roosevelt Road, Taipei 10617, Taiwan\\
$^{3}$Graduate Institute of Astronomy, National Central University, Zhongli,Taoyuan 32001, Taiwan\\
$^4$CAS Key Laboratory of FAST, National Astronomical Observatories, Chinese Academy of Sciences, Peoples Republic of China\\
$^5$ University of Chinese Academy of Sciences, Beijing 100049, People’s Republic of China\\
$^6$ Institute of Astronomy, National Tsing Hua University, No. 101, Section 2, Kuang-Fu Road, Hsinchu 30013, Taiwan\\
$^7$ Department of Astronomy and Astrophysics, Tata Institute of Fundamental Research, Homi Bhabha Road, Mumbai 400 005, India
}

\date{Accepted 2022 July 5. Received 2022 July 5; in original form 2022 May 6}

\pubyear{2022}

\begin{document}
\label{firstpage}
\pagerange{\pageref{firstpage}--\pageref{lastpage}}
\maketitle

\begin{abstract}
Dust polarization observations at optical wavelengths help to understand the dust grain properties and trace the plane-of-the-sky component of the magnetic field.  In this study, we make use of the $I$-band polarization data acquired from AIMPOL  along with the  distances ($d$)  and extinction ($A_{V}$) data to study the variation of polarization fraction ($P$) as a function of $A_{V}$ and $d$ towards the star-forming region, NGC 1893. We employ a broken power-law fit and Bayesian analysis on extinction ($A_{V}$) {\it versus} polarization efficiency ($P$/$A_{V}$) and distance  ($d$) {\it versus} rate of polarization ($P$/$d$). We find that $P$/$A_{V}$ shows a break at an extinction of $\sim$0.9 mag, whereas  $P/d$ exhibits a break at a distance of $\sim$1.5 kpc. Based on these, we categorize the dust towards NGC 1893 into two populations:  (i) foreground dust confined to $A_{V}$ $<$ $\sim$1 mag and  distance up to  $\sim$2 kpc and (ii) Perseus spiral arm dust  towards NGC\,1893 characterized with $A_{V}$ $>$ $\sim$1 mag and distance beyond $\sim$2 kpc.  Foreground dust exhibits higher polarization efficiency but a lower polarization rate, whereas Perseus dust shows a lower polarization efficiency but a slightly higher polarization rate. Hence, we suggest that while  polarization efficiency reveals the dust grain alignment, the rate of polarization infers about the distribution of dust grains towards  NGC 1893. Further, we also shed a light on the spatial variation of intrinsic polarization and magnetic field orientation, and other parameters within the intra-cluster medium of NGC\,1893.
\end{abstract}

\quad

\section{Introduction}

Interstellar dust polarization has been studied extensively for more than half a century since it was observed for the first time at visible wavelengths by \citet{Halletal1949} and \citet{Hiltneretal1949a}. It has been proposed that the observed polarization was caused by the dichroic extinction of asymmetrical dust grains, which are aligned with respect to the Galactic magnetic field \citep{Hiltneretal1949b,Davisetal1951}. 

Since magnetic alignment of the dust grains is a prerequisite for the interstellar polarization to occur, several mechanisms have been proposed to explain it \citep[see,][]{Andersson2015}.  Nevertheless, most of the theoretical and experimental evidence  points towards the radiative alignment torque (RAT) mechanism as the most probable mechanism for dust grain alignment,  which can then explain the observed optical/NIR and far-infrared/millimeter polarization in the interstellar medium  \citep{Lazarianetal2007, Lazarian&Hoangetetal2007a, Anderssetal2011}.  According to the RAT mechanism,  when the radiation incidents upon an irregular dust grain, there is a difference in the  extinction cross-sections experienced by the left circular and right circular polarization components of the incident light.  This results in the generation of radiative torques which act on the grains, causing them to spin up due to  increased angular momentum.  In the presence of an external magnetic field,  the grains Larmor precesses around the magnetic field, and the radiative torques acting over the precessing period align the grains with the field \citep{Dolginov1976,Draine&Weingartneretal1996, draine1997}.

When the  background starlight passes through these magnetically aligned dust grains, its electric vector parallel to the long axis of the dust grains encounter a maximum extinction cross-section as compared to that parallel to the short axis.  This differential extinction cross-section causes the transmitted starlight through the dust grains to be partially plane-polarized up to a few percent. The observed polarized light yield,  two important parameters~--~ (i) fraction of polarization ($P$),  which reveals the averaged dust grain properties along the line of sight (LOS), and  (ii) polarization angle ($PA$), which reveals the LOS averaged magnetic field  projected on the plane of the sky (POS) \citep{Davisetal1951,Lazarian&Hoangetetal2007a}.  

It has been observed that in  optical $V$-band,  the observed polarization, $P$, increases with total extinction ($A_{V}$), and the distribution  for various lines of sight suggest,  an upper limit of $P/A_{V}$ < 3\% mag$^{-1}$ \citep{serkowskietal1975}.  A key conclusion from these data is that $P$ does not increase in the same proportion as $A_{V}$ does, owing to several factors  that include (i) the properties of  dust grains themselves such as the  grain size  (e.g, $\sim$0.1~--~0.3 \micron), shape (e.g, oblate or prolate), composition (e.g, silicate, graphite or graphite-silicate),  mantle to core volume ratio,  presence or absence of surface coatings,  (ii) magnetic field orientation weighted according to the distribution of dust grains along the line of sight, (iii) the efficiency of  grain alignment, and (iv) inclination angle between the actual magnetic field with respect to LOS, and (v) the physical conditions of the environment in which dust exist \citep[see][]{Lee&Draineetal1985}. Therefore, observed polarization measurements need to be critically analysed before interpreting the data. 

One of the key concerns is to know whether the polarization measurements of a distant target would reveal the dust grain properties and magnetic field orientation of foreground dust only,  but not of intra-cluster medium of a distant cluster, or all the dust grains along the  line of sight to the target. To be specific about what regime of $A_{V}$ is responsible for the observed polarization observations, one of the promising approaches would be analysing the variation of the ratio $P/A_{V}$, the dust grain polarization efficiency, as a function of $A_{V}$,  e.g,  by means of a power-law fit of the form $P/A_{V} \propto A_{V}^{-\alpha}$.  The power-law index $\alpha$ at different ranges of $A_{V}$ hints at dust grain alignment efficiency at those corresponding layers of dust.  For example, $\alpha = 1$ corresponds to the total loss of dust grain alignment, and $\alpha = 0$ corresponds to the efficiently aligned dust grains at all optical depths \citep[e.g,][]{Pattleetal2019}.  These interpretations are similar to those based on the index $\alpha$  in  the relation $P^\prime \propto {I^\prime}^{-\alpha}$, where $P^\prime$ and $I^\prime$ are the dust emission polarization and intensity, respectively at sub-millimeter wavelength
\citep[e.g,][]{Jonesetal2015,Wangetal2019,Pattleetal2019}.  The study of polarization efficiency  at various wavelengths has been the subject of research  \citep[][]{goodman1992,goodman1995,gerakines1995,anderssonpotter2005,franco2010,WangJWetal2017}.  The majority of these studies have shown that $P/A_{V}$ varies as a function of $A_{V}$ with the $\alpha$ values ranging from  about $-$1 to $-$ 0.5.  

In this work,  we examine the relation between $A_{V}$ {\it versus} $P/A_{V}$ towards the direction of a  prominent star forming region NGC\,1893,  located at a distance of 3.2 kpc \citep[][]{kharchenko}, with an average age of 2$-$3 Myr \citep[][]{masseyphilip1995b}. NGC\, 1893  is an active star forming region hosting young massive O/B type stars and two elephant trunk-like nebulae Sim 129 and 130 shown in Figure \ref{obs_fields}.  Therefore, it is an interesting region to study the dust properties and magnetic field orientations influencing star formation and polarization.  This region has been investigated extensively over several years \citep{gaze1952,masseyphilip1995b,marcnegu2002,Maheswaretal2007,sourabhsharma2007} and has existing polarization data \citep{Eswaraiahetal2011}.

Another critical parameter that can be used to study the dust  properties is  the ratio of polarization to distance, $P/d$,  as a function of distance ($d$) along a particular LOS. $P/d$ is known as the rate of polarization. This relation is analogous to the rate of extinction,  $A_{V}/d$,   that is,  the ratio of visual extinction ($A_{V}$) to distance ($d$),  the mean  of which is found to be $\approx$ 1.8 mag kpc$^{-1}$ \citep[see,][]{whittet2018}.  Recently, several researchers \citep[e.g.,][]{Eswaraiahetal2011,brajeshkumar2014,  WangJWetal2017, leecd2018, eswaraiah2019, brajeshandeswar2019}    have studied the variation of  $P/d$ using the polarization data from \citet{Heilesetal} and distances from Hipparchus catalogue \citep{hipparchus}  to show that $P$  increases as a function of distance towards several targets in the solar neighbourhood. A common limitation of all such studies have been that they are confined to the solar neighbourhood or to  particular distances where the clusters are located.  In this context, studying $P/d$ as a function of distance towards NGC\,1893, being at a long distance  beyond the solar neighbourhood,  can  shed light on the distribution of polarizing dust grains and presence of dust clouds along the LOS.  It can also help to distinguish the foreground dust from that in the intra-cluster medium by revealing the distance at which the $P/d$ shows a change.

\citet{Eswaraiahetal2011} have observed a constant magnetic field orientation  beyond 1 kpc along the direction of NGC 1893.  It was uncertain whether this is due to the uniform and dominant magnetic field in the foreground dust medium, or a truly constant  field  all up to the distance of NGC 1893.  The revised distances determined from the recent {\it Gaia} EDR3 observations for both foreground and background  fields and cluster members  plus the polarization data  provides an  opportunity to study the variation of polarization and magnetic fields as a function of distance for a statistically significant sample of stars.  It is possible to correct for the foreground contribution in the stokes vector form and therefore to  study the variation in the  dust properties and magnetic fields within a distant cluster.   In this work, in addition to  $A_{V}$ {\it versus} $P/A_{V}$, we also examine $d$ {\it versus} $P/d$   utilizing the newly acquired polarization data along with the existing data \citep{Eswaraiahetal2011},  stellar distances from {\it Gaia} EDR3 \citep{Bailer-JonesRybizkiFouesneauetal2021}, and the extinction values from Bayestar19 three-dimensional (3D) dust map \citep{Greenetal2019}. 

Section \ref{section2} describes our new observations,  and data reduction,  in addition to  archival data. Section \ref{section3} presents the analyses. 
Results are given in Section \ref{section4},   including the fit  with the broken power-law and Bayesian analyses on  $A_{V}$~--~$P/A_{V}$   and  $d$~--~$P/d$ plots, foreground subtraction,  magnetic field morphology in NGC 1893, and the variation of various parameters of the confirmed cluster members as a function of radial distance, etc. We discuss our results in Section \ref{section5}. Finally, Section \ref{section6} summarizes our findings and gives the conclusions.

\section{Observations and data reduction and archival data}\label{section2}

\subsection{Polarization data from ARIES IMaging POLarimeter (AIMPOL)}\label{2.1}

\begin{figure*}
\centering
\resizebox{15.2cm}{13.8cm}{\includegraphics{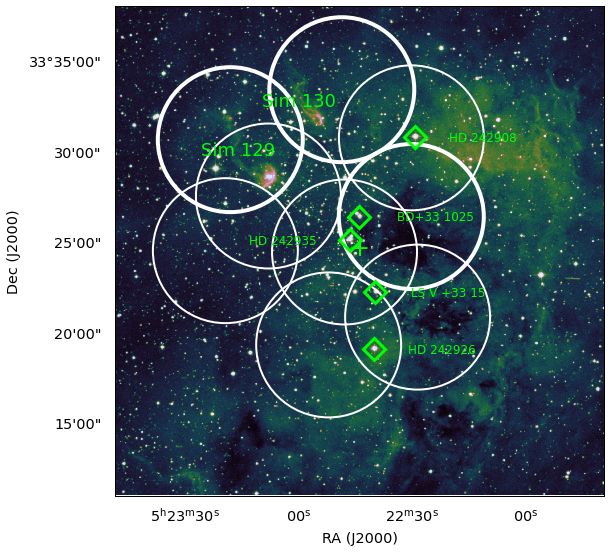}}
\caption{Colour composite image made using WISE 12 $\mu$m (red), 2MASS K-band (green), and DSS R-band (blue) images of the star-forming region NGC\,1893.  Circles each with an 8$\arcmin$ diameter,  denote the
observations acquired with AIMPOL in 2008 (six thin circles) and 2013 (three thick circles).  Cluster centre $(RA,Dec)_{J2000}$=(05$^{h}$:22$^{m}$:44$^{s}$, +33$\degr$:24$\arcmin$:42$\arcsec$) is marked with a green cross.  Five O-type stars HD\,242935, HD\,242926, HD\,242908, BD+33$\degr$ 1025 and LS V +33$\degr$ 15 are shown with green diamond symbols. The location of two emission nebulae Sim 129 and Sim 130 are also shown.
\label{obs_fields}}
\end{figure*}

\begin{figure}
\centering
\resizebox{8.5cm}{15.5cm}{\includegraphics{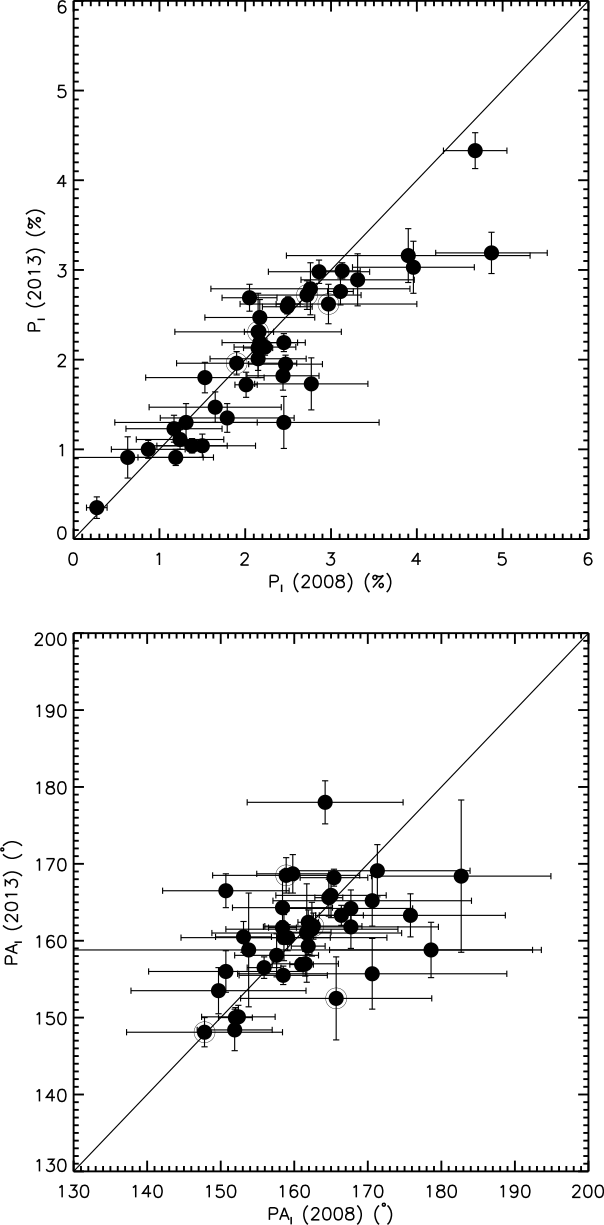}}
\caption{{\it Top:} $P_{{I}_{2008}}$ versus $P_{{I}_{2013}}$ of 40 common stars (filled circles). {\it Bottom:}
$PA_{{I}_{2008}}$ versus $PA_{{I}_{2013}}$ (filled circles). The stars shown with encircled filled circles are the two H$\alpha$ sources and
two Class II/Class III sources. The measurements of 2008 have relatively larger error bars than that of 2013. A thick straight line is drawn with a slope 1 ($=45\degr$) in both
panels to check the one-to-one correlation between the polarization measurements at two epochs.
\label{PPA20082013}}
\end{figure}

Deeper $I$-band polarimetry observations of three fields towards  NGC 1893  were carried out using the ARIES Imaging Polarimeter (AIMPOL) \citep[ ][]{Rautelaetal2004} on 04 November 2013,   mounted on the Cassegrain focus of the 104 cm Sampurnanand Telescope at ARIES in Nainital, India.  In addition, we also  reanalysed the six fields of $I$-band data towards NGC\, 1893 acquired  on the two nights of 08 and 09 November 2008.  The detailed reduction and analysis procedures of AIMPOL data can be found in \citet{Eswaraiahetal2011,Eswaraiahetal2012,Eswaraiahetal2013}.  Figure \ref{obs_fields} depicts the field around NGC 1893. The six fields observed in 2008 and three fields in 2013 are indicated.  Instrumental polarization of AIMPOL has been consistently estimated to be less than 0.1$\%$ \citep[cf. ][]{Eswaraiahetal2011,Eswaraiahetal2012}.  The offset in the polarization angle is determined by observing  the polarized standard star HD\,25443 on 04 November 2013 for which the polarization and polarization angle in $I$-band are estimated  4.16$\pm$0.10$\%$ and 134$\degr\pm$1$\degr$, respectively, consistent with the standard polarization 4.25$\pm$0.04$\%$ and polarization angle 134.2$\degr\pm$0.3$\degr$ in $I$-band  for HD\,25443 \citep{Schmidtetal1992}. All the polarization measurements, acquired on 04 November 2013, are corrected for the instrumental polarization and offset in the polarization angle.  We found 40 common stars observed  in both years of  2008 and 2013. The values of polarization ({\it Top}) and polarization angles ({\it Bottom}) measured in 2008 and 2013 are found to match closely with each other, as shown in Fig. \ref{PPA20082013}. The stars shown with encircled filled circles are the two H$\alpha$ sources \citep{Maheswaretal2007} and two Class II/Class III sources \citep{Pandeyetal2014}. The measurements made in 2013 have smaller uncertainties owing to longer exposure time than those measured in 2008.  Therefore, we adopt the measurements of 2013 for common sources in both epochs,  and for the sources otherwise detected in only one epoch,  we use the corresponding epoch data in  our analysis.

\begin{figure*}
\resizebox{12cm}{14cm}{\includegraphics{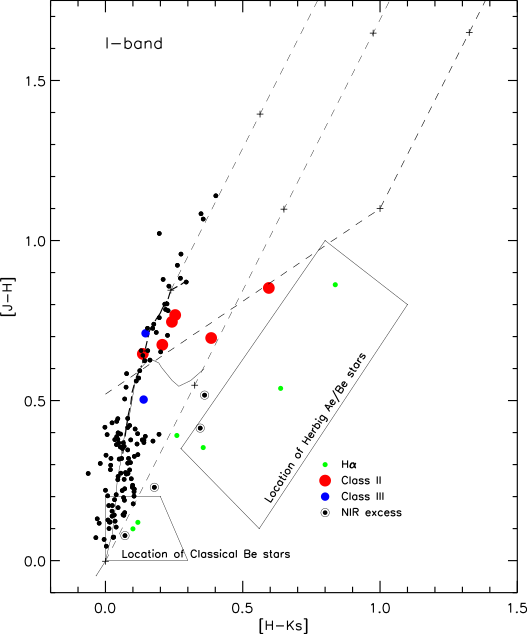}}
\caption{ [J-H] vs [H-K$_{\rmn s}$] colour-colour diagram for 173 stars with I-band polarimetry.
Red and blue-filled circles are Class II and Class III sources \citep{Pandeyetal2014}.
Small green-filled circles are the H$\alpha$ sources \citep{Maheswaretal2007}. Other NIR-excess sources were represented with encircled filled circles. The rest of the stars were shown with filled black circles.
The solid and thick dashed curves represent the unreddened MS and giant branch \citep{BessellBrett1988}, respectively.  The dashed line indicates the locus of unreddened CTTSs \citep{MeyerCalvetHillenbrand1997}.  The parallel dashed lines are the reddening vectors drawn from the tip (spectral type M4)
of the giant branch (``upper reddening line"), from the base (spectral type $A0$) of the MS branch
(``middle reddening line") and from the tip of the intrinsic CTTS line (``lower reddening line").
The location of Be stars \citep[cf.][]{Doughertyetal1994} and the
location of Herbig Ae/Be stars \citep[cf.][]{Hernandezetal2005} are also shown.
\label{NIR_IJH}}
\end{figure*}

Before we perform the analyses, we exclude the stars with possible intrinsic polarization based on the near-infrared (NIR) colour-colour diagram.   Here,  we assume that the intrinsic polarization can only be caused by the stars with circumstellar disks.  Figure \ref{NIR_IJH} shows the [J-H] vs [H-K$_{\rmn s}$] NIR colour-colour diagrams for 173 stars  in $I$-band whose signal-to-noise ratio in polarization, $P/\sigma_{P}$, is greater than 3,  i.e, possibly with intrinsic polarization.  The $JHK_{s}$ photometric data have been obtained from
2MASS Point Source Catalogue \citep{Cutrietal2003}.  Red and blue-filled circles are the Class II and Class III sources \citep{Pandeyetal2014}.    
Small green-filled circles are the H$\alpha$ sources \citep{Maheswaretal2007}. Other NIR-excess sources  are represented with encircled filled circles.  In total, 21 stars,  as mentioned above, may have intrinsic polarization. Therefore, out of 173 stars,  we  utilize the polarization data of 152 stars for the analyses.   Figure \ref{NIR_IJH} shows that most stars are not highly extincted, but rather are  distributed around the main sequence (MS) locus,
with a few  along the giant locus.

\subsection{Distances from GAIA Early Data Release 3}

We obtain the distances  from \citet{Bailer-JonesRybizkiFouesneauetal2021}.  They have estimated distances using stellar parallaxes of 1.47 billion stars and have published  {\it Gaia} Early Data Release 3 (hereafter {\it Gaia} ERD3). Even though {\it Gaia} ERD3 have a significant improvement in both precision and accuracy of astrometry and broad-band photometry, the majority of these stars are  distant or faint that their  parallax uncertainties are large, thereby precluding a simple inversion of stellar parallax to provide a distance. Therefore, they use a probabilistic approach to estimate the  stellar distances using a prior constructed from a three-dimensional model of our galaxy,   which includes the interstellar extinction and Gaia’s variable magnitude limit.  Using this method,  Bailer-Jones estimate two types of distances: (i) Geometric and (ii) Photogeometric.  The first one uses the parallax together with a direction-dependent prior on distance, whereas the second one additionally uses the colour and apparent magnitude of a star, by exploiting the fact that stars of a given colour have a restricted range of probable absolute magnitudes and extinction.  There exist geometric distances for 1.47 billion stars, and 92\% of these stars have photogeometric distances.
Tests on simulated data and external validation show that the photogeometric estimates generally have higher accuracy and precision for fainter and distant stars \citep[][]{Bailer-JonesRybizkiFouesneauetal2021}. Therefore, we utilize photogeometric distances for our study.  We cross-match the coordinates of 152 stars having polarization data with the Bailer-Jones catalogue towards NGC\,1893. Within  a 0.5 $\arcsec$ matching radius, we find  151 stars have photogeometric distances. They are given in Table \ref{sampletable}.

\subsection{Extinction data}
We use the total extinction at V-band, $A_{V}$, data for our analyses. $A_{V}$ data are extracted from \citet{Greenetal2019}. They use stellar photometry from optical Pan-STARRS 1 and near-infrared 2MASS  photometry along with the {\it Gaia} DR2 parallaxes to infer distances and reddening to $\sim$800 million stars. The reddening values of these stars are traced along different lines of sight, based on which they build a map of dust reddening in three-dimensions \citep{Greenetal2019} referred to as Bayestar19. Extinction towards a particular line of sight but at different distances can be extracted using  the {\sc Python} based module, dust maps\footnote{\url{{https://dustmaps.readthedocs.io/en/latest/modules.html##module-dustmaps.bayestar}}}. The extinction map is structured as a set of sight lines, each of which contains multiple samples of cumulative dust reddening as a function of distance. Each sight line is identified by a HEALPix nside parameter and a nested pixel index. Within each sight line, cumulative reddening is given at discrete distances, spaced evenly in the distance modulus. Thus, for each pixel, it provides multiple samples from the posterior of dust reddening. It also gives quality assurance information for each pixel, i.e., whether the fit converges in the pixel and the minimum/maximum reliable distance moduli in the pixel
\citep[][]{Greenetal2019}\footnote{\url{{https://dataverse.harvard.edu/dataset.xhtml?persistentId=doi:10.7910/DVN/2EJ9TX}}}.

We use the above-mentioned {\sc Python} package `dust maps' which provides functions for querying and downloading the Bayestar19 3D dust map.  The reddening values from the Bayestar19 dust map are obtained by providing the  equatorial coordinates of stars and their distances as  input. Thus, the reddening values corresponding to the maximum  posterior density at a particular coordinate and distance are returned as output.  Finally, the obtained reddening values are checked for their quality by examining whether the line-of-sight fit has converged in a given pixel and also whether the input distance is reliable within the range of the considered distances.  Hence,  we find the reddening values for 151 stars.  Of which, 8 do not have reliable distance information, although their line-of-sight fits are converged.  These eight stars are excluded from further analyses.  Thus, the obtained reddening values for 143 stars are converted into extinctions  using the following relations \citep{Greenetal2019} \footnote{\url{{http://argonaut.skymaps.info/usage##units}}}:
\begin{eqnarray}
E(g-r)=0.901\cdot E_{\mathrm{bayestar19}} \nonumber\\
E(B-V)=0.981\cdot E(g-r)\nonumber\\
A_{V}=R_{V} \cdot E(B-V)
\end{eqnarray}
where $E_{\rmn{bayestar19}}$ is the reddening value obtained from Bayestar19 3D dust map.  The $E(g-r)$ and $E(B-V)$ are the reddening values corresponding to $g-r$ and $B-V$ colours, $A_{V}$ is the total extinction in the visual band, and $R_{V}$ is the total-to-selective extinction of 3.1 \citep{Savage&Mathisetal1979,Cardellietal1989}.   Since the dustmaps module does not provide uncertainties in reddening values, they are extracted from Galactic Interstellar Extinction (GALExtin\footnote{\url{www.galextin.org}}; \citealt{Amoresetal2021}).

The maximum distance up to which the Bayestar19 3-D extinction map reddening values are reliable are estimated based on the distance at which the main sequence stars disappear along a particular LOS \citep[][]{Greenetal2019}.  Since the stellar cluster NGC 1893 which is the subject of our study has a galactic latitude of -1.6342\textdegree, and is located close to the galactic midplane,   the stellar number density in this region  is significantly high \citep[see,  Figure 6 (a) of][]{sourabhsharma2007}{}.   So, there won't be any dearth of main sequence stars at a particular distance along any LOS in this region. So the extinction ($A_{V}$) values we have extracted are valid up to a maximum of distance of 7 kpc which is the maximum photogeometric distance that we have used in our study. 

To check  the reliability of our  reddening values  obtained from Bayestar19 dust map \citep[][]{Greenetal2019} , we compare our reddening values with those from Bayestar15 \citep[][]{bayestar2015}, Bayestar17 \citep[][]{bayestar17} and Chen2019 \citep[][]{Chen2019} dust maps.   Bayestar15 is a 3D  map of dust reddening using  Pan-STARRSS 1 and 2MASS photometry based on a probabilistic approach,  covering over a three-quarters of the sky. Bayestar17 is an improved version of Bayestar15 with the inclusion of additional Pan-STARRS 1 data, updated methodology and finer angular resolutions at high extinction regions. Chen2019 is another all sky map of dust reddening in 3D computed using the parallax and optical photometry from {\it Gaia} DR2 and infrared photometry from 2MASS and WISE.  We find that most of our reddening values have a one to one correspondence with the reddening values obtained from these three dust maps, thereby attesting that our reddening values are reliable, and these $A_{V}$ data are also given in Table \ref{sampletable}.

\subsection{Dust temperature using IRAS data}
One of the key predictions of RAT theory is the positive correlation between the degree of polarization and dust temperature \citep[][]{Lazarian&Hoangetetal2007a,Andersson2015}. In order to examine whether this relation exists in NGC\,1893 region,  we extract the dust temperature map. Since  the current target does not have {\it Herschel} data, we use Infrared Astronomical Satellite (IRAS) maps at 60 and 100 $\mu$m taken from SkyView Virtual observatory\footnote{\url{https://skyview.gsfc.nasa.gov/current/cgi/titlepage.pl}}.  The dust temperature in each pixel of the IRAS image is obtained by assuming that the dust in a single beam is isothermal and that the observed ratio of 60 to 100 $\mu$m emission follows a blackbody radiation from dust grains at temperature $T_{d}$, modified by a power-law emissivity spectral index $\beta$.  We use the following relation to derive $T_{d}$ \citep{Schneeetal2005}:
\begin{equation}
    R = 0.6^{-(3+\beta)}\cdot\frac{e^{144/T_{d}} -1}{e^{240/T_{d}} -1}
\end{equation}
where $R$ is the ratio of 60 to \SI{100}{\micro\meter} flux densities, $\beta$ is the dust emissivity spectral index which is approximately taken as 2,  and $T_{d}$ is the dust temperature at each pixel of the IRAS image.  We cross-match the list of cluster stars having polarization measurements with the pixels of $T_{d}$ map. Since the resolution of IRAS 60/100 $\mu$m, and hence the $T_{d}$ map, is 2\arcmin, we estimate mean and standard deviation in $T_{d}$ over the pixels within 2\arcmin \hspace{0.08cm} diameter around each star having polarization data.

\begin{table*}
\caption{Sample data table containing the year of observation, equatorial coordinates, distance, polarization data in I-band, 2MASS photometric data, optical extinction and temperature of the stars that we have used in this study.}
\label{sampletable}

\begin{tabular}{lcccccccl}

 \hline

 SNo &  Year & RA (J2000) & Dec (J2000) & Distance $\pm$ $\sigma$ &  P $\pm$ $\sigma$  &   PA $\pm$ $\sigma$      \\
 \hline

  &       &   (degree) & (degree)    &           (kpc)         &         (\%)       &        (degree)\\ 
 \hline
 1  &  2008 &  80.665346 &  33.371732 &  3.00 $\pm$ 0.16 &  2.12 $\pm$  0.18 &  151.0 $\pm$ 2.0\\
2  &  2008 &  80.723805 &  33.392084 &  2.22 $\pm$ 0.10 &  2.15 $\pm$  0.48 &  167.0 $\pm$ 6.0\\
3  &  2008 &  80.719046 &  33.386870 &  2.32 $\pm$ 0.12 &  2.37 $\pm$  0.49 &  164.0 $\pm$ 6.0\\
4  &  2008 &  80.717592 &  33.384285 &  2.35 $\pm$ 0.15 &  2.54 $\pm$  0.31 &  162.0 $\pm$ 4.0\\
5  &  2008 &  80.688602 &  33.371345 &  0.86 $\pm$ 0.01 &  1.46 $\pm$  0.23 &  174.0 $\pm$ 4.0\\
6  &  2008 &  80.699181 &  33.368585 &  0.94 $\pm$ 0.02 &  1.65 $\pm$  0.38 &  175.0 $\pm$ 7.0\\
7  &  2008 &  80.741272 &  33.368718 &  2.68 $\pm$ 0.78 &  2.79 $\pm$  0.49 &  158.0 $\pm$ 5.0\\
8  &  2008 &  80.771703 &  33.359300 &  2.87 $\pm$ 0.16 &  2.23 $\pm$  0.63 &  156.0 $\pm$ 8.0\\
9  &  2008 &  80.622089 &  33.514012 &  3.01 $\pm$ 0.21 &  2.86 $\pm$  0.12 &  147.0 $\pm$ 1.0\\
10  &  2008 &  80.699321 &  33.476143 &  0.84 $\pm$ 0.01 &  1.88 $\pm$  0.21 &  165.0 $\pm$ 3.0\\
11 &  2008 &  80.751907 &  33.496616 &  2.51 $\pm$ 0.12 &  2.79 $\pm$  0.27 &  161.0 $\pm$ 3.0\\
12 &  2008 &  80.731302 &  33.487467 &  3.10 $\pm$ 0.25 &  3.39 $\pm$  0.58 &  159.0 $\pm$ 5.0\\
13 &  2008 &  80.767385 &  33.470600 &  2.92 $\pm$ 0.17 &  3.12 $\pm$  0.48 &  155.0 $\pm$ 4.0\\
14 &  2008 &  80.725975 &  33.445086 &  2.81 $\pm$ 0.16 &  3.00 $\pm$  0.59 &  154.0 $\pm$ 6.0\\
15 &  2008 &  80.782324 &  33.467270 &  2.14 $\pm$ 0.12 &  2.82 $\pm$ 0.54  &  159.0 $\pm$ 6.0\\
16 &  2008 &  80.762063 &  33.454518 &  1.63 $\pm$ 0.08 &  2.25 $\pm$  0.69 &  159.0 $\pm$ 9.0\\
17 &  2008 &  80.741849 &  33.443460 &  1.47 $\pm$ 0.50 &  2.36 $\pm$  0.26 &  160.0 $\pm$ 3.0\\
18 &  2008 &  80.713246 &  33.426608 &  3.29 $\pm$ 0.28 &  3.03 $\pm$  0.73 &  150.0 $\pm$ 7.0\\
19 &  2008 &  80.744290 &  33.400895 &  2.73 $\pm$ 0.22 &  3.47 $\pm$ 0.52 &  160.0  $\pm$ 4.0\\
20 &  2008 &  80.765901 &  33.400821 &  3.28 $\pm$ 0.20 &  2.90 $\pm$  0.46 &  161.0 $\pm$ 5.0 \\
%\bottomrule
\end{tabular}
\hspace{5cm}
%\scalebox{1}{
\begin{tabular}{lccccccl}
 \hline

 SNo & J $\pm$ $\sigma$ &   H $\pm$ $\sigma$ & K $\pm$ $\sigma$ & $A_{V}$ $\pm$ $\sigma$ &  Temperature $\pm$ $\sigma$ \\
 \hline
 
    &         (mag)    &         (mag)      &      (mag)     &      (mag)         &             (K)  \\
 \hline
 1  &  9.00$\pm$ 0.02 &        8.89 $\pm$          0.02 &        8.79 $\pm$           0.02 &     1.67 $\pm$        0.12 &           28.72 $\pm$        0.77 \\
2  &       11.58 $\pm$           0.02 &       11.57 $\pm$           0.02 &       11.54 $\pm$           0.02 &     0.79 $\pm$        0.02 &               -\\
3  &       11.82 $\pm$          0.02 &       11.76 $\pm$           0.02 &       11.71 $\pm$           0.02 &     0.79 $\pm$        0.02 &               - \\
4  &       11.02 $\pm$           0.02 &       10.99 $\pm$           0.02 &       10.99 $\pm$           0.02 &     0.79 $\pm$        0.02 &               -  \\
5  &       10.13 $\pm$           0.03 &       10.06 $\pm$           0.02 &        9.99 $\pm$           0.02 &     0.63 $\pm$         0.1 &               -  \\
6  &       10.92 $\pm$          0.02 &       10.58 $\pm$           0.02 &       10.49 $\pm$           0.02 &     0.63 $\pm$         0.1 &               -  \\
7  &       11.91 $\pm$           0.02 &       11.82 $\pm$           0.02 &       11.81 $\pm$           0.02 &     1.07 $\pm$        0.09 &               - \\
8  &       11.44 $\pm$           0.02 &       10.70 $\pm$           0.02 &       10.50 $\pm$           0.02 &     1.24 $\pm$        0.11 &               -  \\
9  &        8.28 $\pm$           0.02 &        8.25 $\pm$           0.03 &        8.18 $\pm$           0.02 &     1.07 $\pm$        0.09 &           30.87 $\pm$        0.31 \\
10 &        8.82 $\pm$          0.02 &        8.22 $\pm$           0.02 &        8.06 $\pm$           0.02 &     0.68 $\pm$        0.02 &               - \\
11 &       11.64          &       11.53 &       11.43 $\pm$           0.03 &     0.79 $\pm$        0.05 &           - \\
12 &       12.48            0.02 &       12.09 $\pm$           0.02 &       11.94 $\pm$           0.02 &     1.15 $\pm$        0.05 &           31.21 $\pm$         0.4 \\
13 &       12.77          &       12.57 $\pm$          0.05 &       12.47 $\pm$           0.04 &     0.88 $\pm$        0.07 &               - \\
14 &       12.80 $\pm$           0.03 &       12.68 $\pm$           0.04 &       12.62 $\pm$           0.03 &     0.87 $\pm$        0.05 &               -  \\
15 &       12.48 $\pm$           0.02 &       12.31 $\pm$           0.03 &       12.24 $\pm$          0.02 &     0.83 $\pm$        0.02 &               - \\
16 &       12.77 $\pm$           0.02 &       12.43 $\pm$           0.02 &       12.34 $\pm$           0.02 &     0.82 $\pm$        0.02 &               -  \\
17 &       11.46 $\pm$           0.02 &       11.32 $\pm$           0.02 &       11.26 $\pm$           0.02 &     0.82 $\pm$        0.03 &               - \\
18 &       13.44 $\pm$           0.03 &       13.33 $\pm$           0.04 &       13.22 $\pm$          0.04 &     2.16 $\pm$        0.04 &           30.58 $\pm$        0.64 \\
19 &       12.40 $\pm$           0.02 &       12.26 $\pm$           0.02 &       12.21 $\pm$           0.02 &     0.86 $\pm$        0.07 &               -  \\
20 &       11.90 $\pm$          0.02 &       11.24 $\pm$           0.02 &       11.05 $\pm$           0.02 &     1.15 $\pm$       0.08 &           31.64 $\pm$        0.53 \\
%\bottomrule
\hline

\end{tabular}
\end{table*}
\section{analyses}\label{section3}

In this section we analyse how the polarization efficiency ($P/A_{V}$) varies as a function of extinction ($A_{V}$)  and (ii) rate of polarization ($P/d$) varies as a function of distance ($d$). We used these analyses to examine whether the observed polarization measurements reveal the dust properties and magnetic field orientation of the foreground medium alone but not of the intra cluster medium, or of the dust grains distributed along the entire line of sight towards NGC\,1893.  After confirming the latter one that our polarization measurements consist of a contribution from the magnetically aligned dust grains in the intra-cluster medium,  we try to quantify the foreground polarization contribution and subtract it from the measurements of cluster members. Eventually, we examine if there are changes in intrinsic polarization parameters, dust temperature, extinction, and polarization efficiency within the cluster region.

We model the plots with two different methods:~--~ (i) broken power-law with a Levenberg-Marquardt non-linear least squares fitting method and (ii) broken power-law but with a Bayesian approach.

\subsection{Broken Power-law fit}\label{subsec:brokenpowerlaw}

\begin{figure}
\captionsetup[subfloat]{labelformat=empty}
\subfloat[\label{fig:pbyavbroken}]{\includegraphics[width=3.35in,height=2.5in]{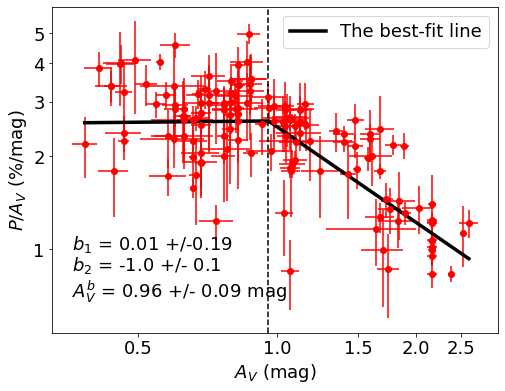}}\hspace{1em}\\
\subfloat[ \label{fig:pbydbroken}]{\includegraphics[width=3.25in,height=2.5in]{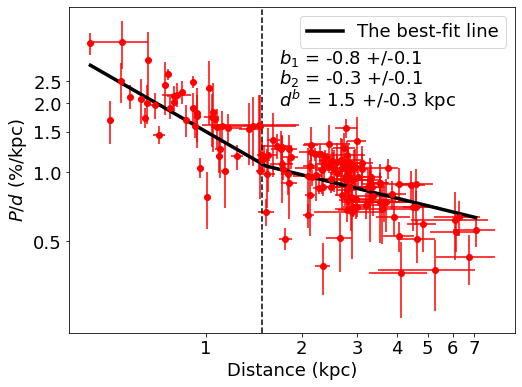}}
\caption{$A_{V}$ {\it versus} $P/A_{V}$ plot ({\it Top}) and $d$ {\it versus} $P/d$ plot ({\it Bottom}). Best fit broken power-law is denoted with thick line. The dotted vertical line corresponds to $A_{V}^b$ and  $d^{b}$ at which the power-law breaks. Best-fit parameters are also overlaid.} 
\label{brokenpowerlawplots}
\end{figure}

The  relation $A_{V}$~--~$P/A_{V}$ is fitted with broken power-law of the form using $scipy-curvefit$ module in {\sc Python}:

\begin{eqnarray}\label{eq:brokenpowerlawav}
    y=\begin{cases}
 a\left(\frac{x}{A_{V}^b}\right)^{b_{1}}, & \text{for}~~ x \leq {A_{V}}^{b}\\[10pt]
a\left(\frac{x}{A_{V}^b}\right)^{b_{2}}, &
 \text{for}~~ x > {A_{V}}^{b},
 \end{cases}
\end{eqnarray}
where $x$ is $A_{V}$,  $A_{V}^b$ is the $A_{V}$ at which the power-law breaks, and $a$ is a coefficient. The $b_{1}$ and $b_{2}$ are power-law indices before and after $A_{V}^b$.  Similarly, the data of $d$~--~$P/d$ are also fitted with the broken power-law of the similar form to equation \ref{eq:brokenpowerlawav}, where $A_{V}$ and $A_{V}^b$ are replaced with distance ($d$) and distance break $d^b$ at which the power-law breaks, respectively.  Figure \ref{brokenpowerlawplots}  show the data and corresponding best fits for $A_{V}$~--~$P/A_{V}$ and $d$~--~$P/d$, respectively. The best-fit parameters are given in Table \ref{tab:brokenpowerlawparams}.

To check whether the broken power-law model of $A_{V}$~--~$P/A_{V}$ and $d$~--~$P/d$ fit the data better than a single power-law,  we perform ${\chi}^2$-difference test and also estimate Akaike Information Criterion (AIC) and Bayesian Information Criterion (BIC) values.  For $A_{V}$~--~$P/A_{V}$ with the difference in the  degrees of freedom  2,  the  ${\chi}^2$-difference is 9.255  which is  statistically significant as the corresponding p-value turns out to be 0.01 which is less than the statistically accepted 0.05 significance level, meaning that the model with more free-parameters or the broken power-law  fits the data better. Furthermore, the AIC and BIC values for the broken power-law  are lower than that of single power-law fit, confirming that the broken power-law is a better model than single power-law for $A_{V}$~--~$P/A_{V}$.  For $d$~--~$P/d$ again with the difference in the degrees of freedom  2, the ${\chi}^2$-difference is 1.032, which is statistically insignificant as the corresponding p-value turns out to be 0.6 which is greater than the significance level of 0.05, implying that both models fit the data equally well. But lower values of AIC and BIC  when compared to that of single power-law suggests that the broken power-law fits the data better.  Therefore, broken power-law model can be accepted for fitting  both $A_{V}$~--~$P/A_{V}$ and  $d$~--~$P/d$ relations.

\begin{table*}
\caption{Best fit parameters, based on the broken power-law, for $A_{V}$ ~--~ $P/A_{V}$ and $d$ ~--~$P/d$}
\label{tab:brokenpowerlawparams}
\begin{tabular}{lcccccl}

 \hline
 
 Relation & $a$ & $b_{1}$ & $b_{2}$ &  $A_{V}^{b}$ or $d^{b}$ \\
 \hline

 $P/A_{V}$~--~$A_{V}$& 2.6 $\pm$ 0.2  & 0.01 $\pm$ 0.19  & -1.0 $\pm$ 0.1 & 0.96 $\pm$ 0.09 mag\\
 \hline

 $P/d$~--~$d$ & 1.1 $\pm$ 0.1 & -0.8 $\pm$ 0.1 & -0.3 $\pm$ 0.1 & 1.5 $\pm$ 0.3 kpc\\
 \hline

\end{tabular}
\end{table*}

\subsection{Bayesian Analyses}\label{subsec:bayesianapproach}

\begin{figure}
\captionsetup[subfloat]{labelformat=empty}
\subfloat[\label{pbyavbayesian}]{\includegraphics[width=3.25in,height=2.5in]{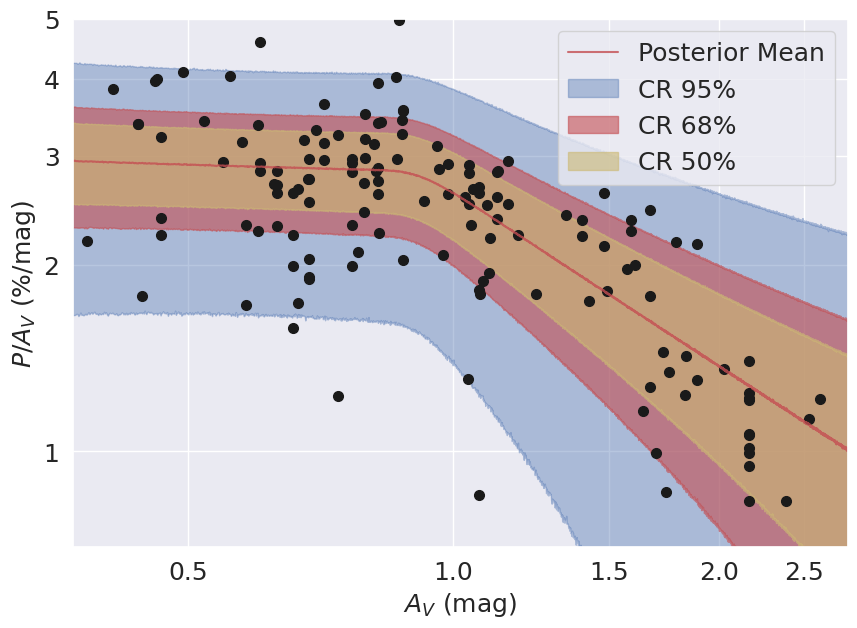}}\hfill\\
\subfloat[\label{pbydbayesian}]{\includegraphics[width=3.25in,height=2.5in,scale=4]{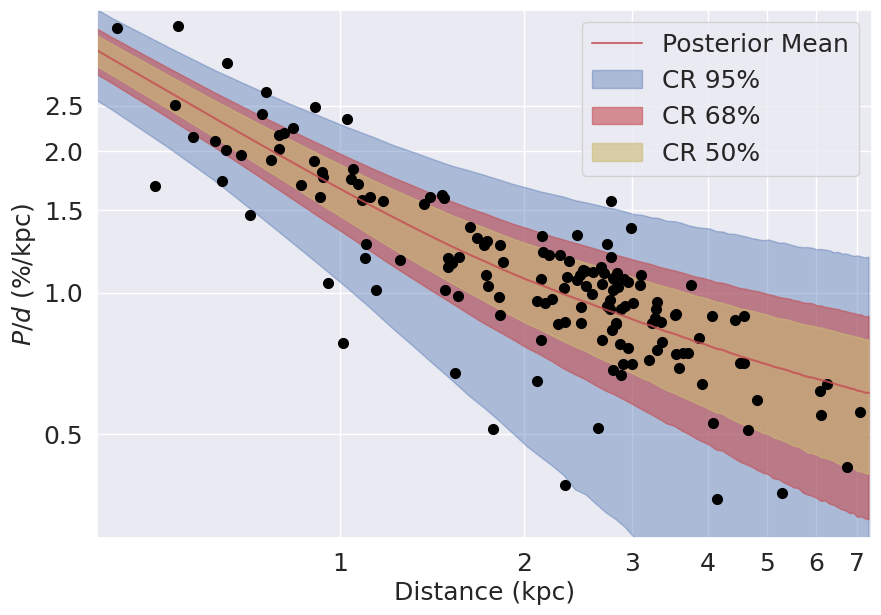}}
\caption{Comparison between the Bayesian posterior prediction and the observations for $A_{V}$ {\it versus} $P/A_{V}$ ({\it Top}) and $d$ {\it versus} $P/d$ ({\it Bottom}). The black line and coloured regions show the mean, 95\%, 68\%, and 50\% confidence regions, predicted by the posteriors as shown in  {\it Top} and {\it Bottom} panels of Figure \ref{bayesianpdf} respectively.}
\label{bayesianplots}
\end{figure}

\begin{figure*}%[htbp]
\captionsetup[subfloat]{labelformat=empty}
\subfloat[ \label{analbayespbyav}]{\includegraphics[width=5.5in,height=4.15in,scale=4]{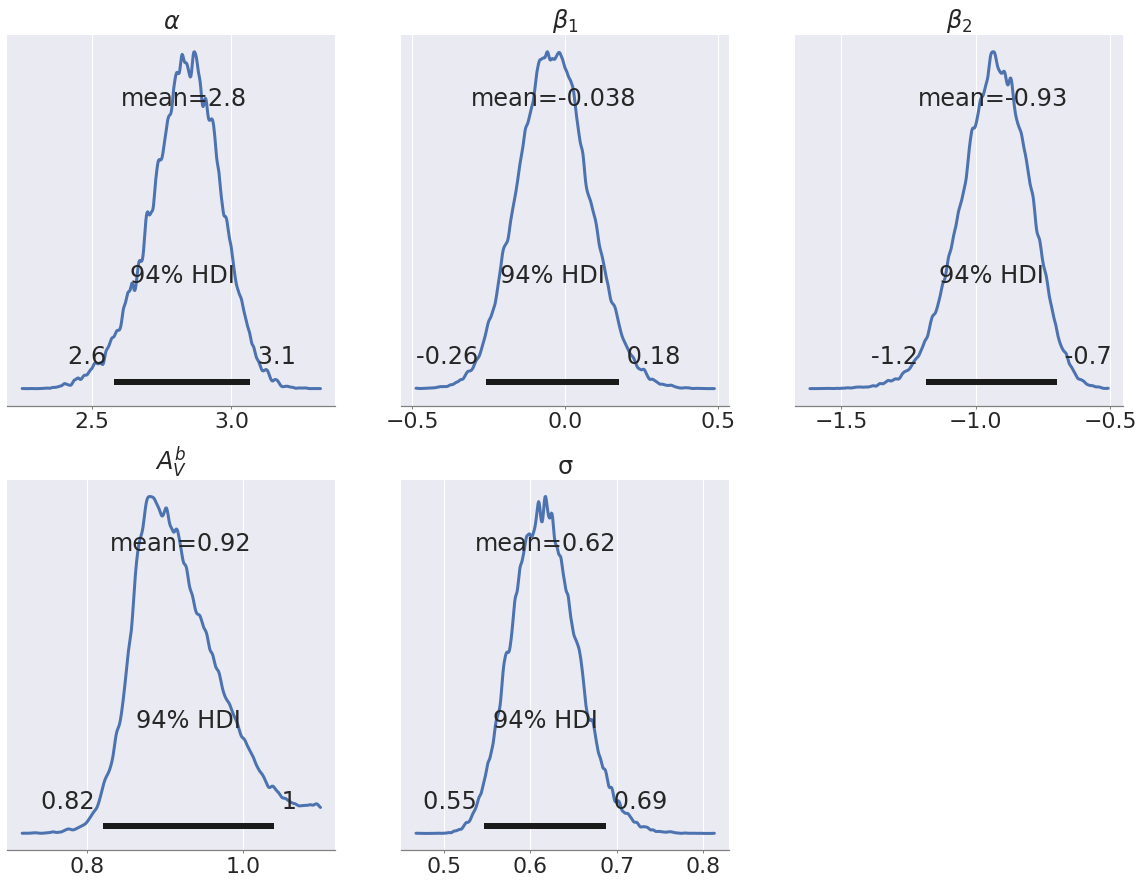}}\hfill\\
\subfloat[ \label{analbayespbyd}]{\includegraphics[width=5.5in,height=4.15in,scale=4]{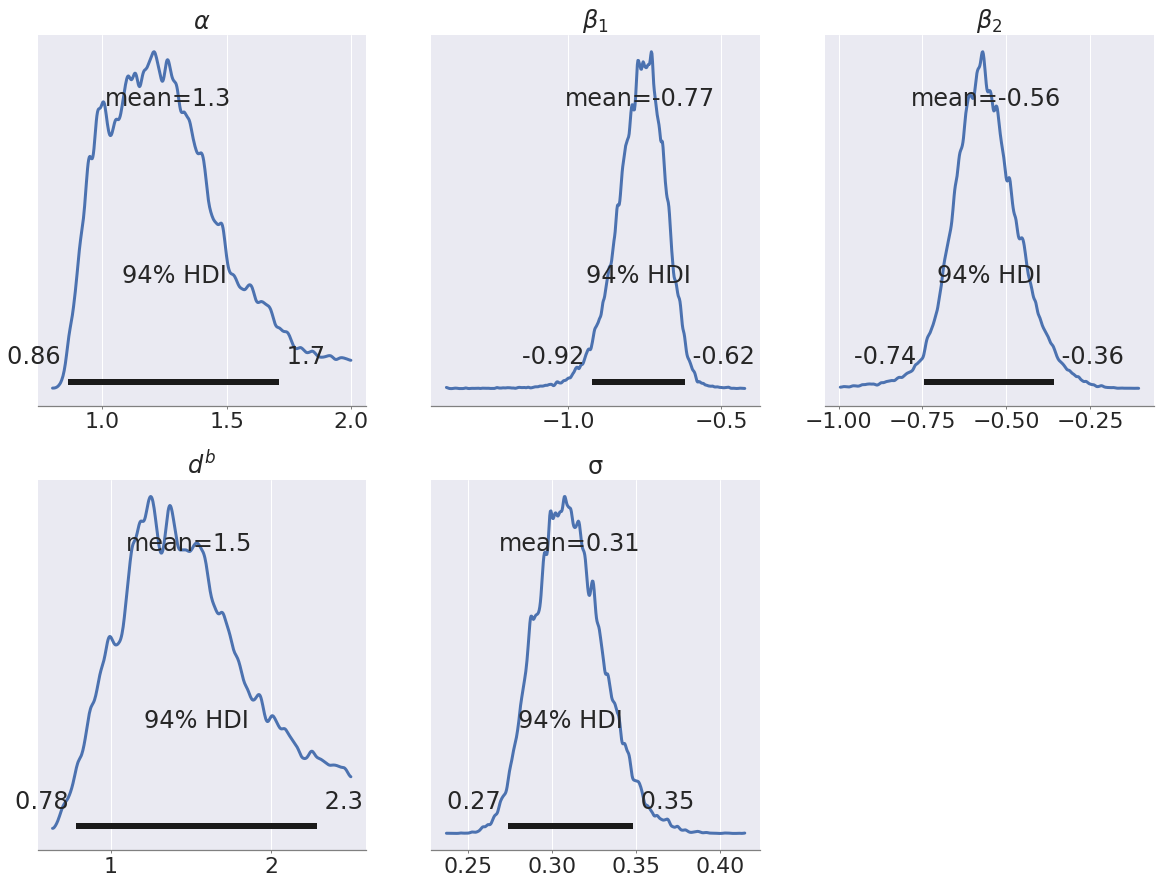}}
\caption{PDF of the model parameters derived using Bayesian model fitting of $A_{V}$ {\it versus} $P/A_{V}$ ({\it Top}) and $d$ {\it versus} $P/d$ ({\it Bottom}) respectively. }
\label{bayesianpdf}
\end{figure*}

We have employed Bayesian analyses to confirm the results obtained from the broken power-law for $A_{V}$~--~$P/A_{V}$ 
and  $d$~--~$P/d$ (above section).  Bayesian analyses can help constrain the posterior distribution of all parameters based on the prior distributions and observational data expressed in the form of a likelihood function. We modeleled the trends followed by $A_{V}$~--~$P/A_{V}$ and  $d$~--~$P/d$ by combining broken power-law with Bayesian approach.  The resulting analyses yield us the posterior mean fit along with the 95\%, 68\%, and 50\%  confidence regions (CRs), and the posterior probability density function (PDF) of each parameter as output as described in appendix \ref{appendixa}.  The 95\%, 68\%, and 50\% confidence regions (CRs) predicted by the posterior distribution for $A_{V}$~--~$P/A_{V}$ and  $d$~--~$P/d$ plots  are shown in Figure \ref{bayesianplots}.  From the figures, it is seen that the majority of the data fall well within 95\% CR predicted by the posterior.  The 94 \% highest density interval (HDI) of each parameter are plotted to represent the data lying within 2$\sigma$ uncertainties. The posterior PDF of the parameters derived from the Bayesian modelling is given in Figure \ref{bayesianpdf}. In Figure \ref{bayesianpdf}, $\alpha$ is a constant, $\beta_{1}$, and $\beta_{2}$ are the power-law indices before and after the break in power-law for both the $A_{V}$~--~$P/A_{V}$.  and  $d$~--~$P/d$ plots. $A_{V}^b$ and $d^b$ are the $A_{V}$ and $d$ values at which the break occurs in $A_{V}$~--~$P/A_{V}$  and  $d$~--~$P/d$ respectively. Similarly, $\sigma$ is the dispersion in the data of $P/A_{V}$ or $P/d$. More details about various parameters are given in the appendix \ref{appendixa}. The parameters constrained through the  Bayesian analyses are reasonably more accurate than those obtained from the broken power-law alone. The constrained results from Bayesian analyses are listed in Table \ref{tab:bayesbrokenpowerlawparams}. 
\begin{table*}
\caption{Best fit parameters based on broken power-law Bayesian model for 
 $A_{V}$~--~$P/A_{V}$ and $d$~--~$P/d$}
\label{tab:bayesbrokenpowerlawparams}
\scalebox{0.8}{
\begin{tabular}{l|c|ccc|ccc|ccc|ccl}

 \hline
 
 Relation  &  $\alpha$ & $\beta_{1}$ &3\%\, HDI ($\beta_{1}$)  & 97 \%\, HDI ($\beta_{1}$)& $\beta_{2}$ & 3\%\, HDI ($\beta_{2}$) & 97 \%\, HDI ($\beta_{2}$)  & $A_{V}^{b}$ or $d^{b}$  & 3\%\, HDI ($A_{V}^{b}$ or $d^{b}$) & 97 \%\, HDI ($A_{V}^{b}$ or $d^{b}$) & $\sigma$ \\
 \hline
 
 $P/A_{V}$~--~$A_{V}$ & 2.8  & 0.04  & -0.26 & 0.18 &-0.9   & -1.2 & -0.7 &  0.9 mag & 0.8 mag & 1 mag & 0.6 \%/mag \\
 \hline

 $P/d$~--~$d$ & 1.3 & -0.8 & -0.9 & -0.6 & -0.6 & -0.7 & -0.4 &  1.5 kpc &0.8 kpc & 2.3 kpc & 0.3 \%/kpc  \\
 \hline

\end{tabular}}

\begin{tablenotes}
\centering
\small
\item HDI: Highest density interval
\end{tablenotes}
\end{table*}


\subsection{ Differential plots in P and PA as a function of distance}\label{3.3}

Stokes parameters ($Q$ and $U$) of all 151 stars are converted from  $P$ and $PA$ (see, section \ref{2.1}) using the relations:

\begin{eqnarray}\label{stokesvalues}
Q = P \cos(2PA)\nonumber \\ 
{\mathrm{and}} \,\,
U = P \sin(2PA).
\end{eqnarray}

To understand the change in polarization parameters in response to the dust layers along the line of sight, we construct differential plots in Stokes parameters  $\Delta Q$~$(= Q_{i+1} - Q_{i}$) and $\Delta U$~$(= U_{i+1} - U_{i}$) {\it versus} distance$_{i+1}$, where $i$ is an immediate foreground star to $(i+1)^{th}$ star.  These $\Delta Q$ and $\Delta U$ are converted into $\Delta P$ and $\Delta PA$ using the relations:
\begin{eqnarray}\label{eq14}
\Delta P&=& \sqrt{\Delta Q^2+\Delta U^2}\nonumber\\
\Delta PA&=&0.5\cdot{arctan}\left(\frac{\Delta U}{\Delta Q}\right)\cdot\left(\frac{180}{\pi}\right)
\end{eqnarray}
 The corresponding errors are given by: 
\begin{eqnarray}\label{eq15}
\sigma_{\Delta P}=\frac{1}{\Delta P}\cdot\sqrt{(\Delta Q\cdot \sigma_{
\Delta Q})^2+ (\Delta U\cdot \sigma_{\Delta U})^2}\nonumber\\
\sigma_{\Delta PA}=\frac{1}{2\Delta P^2}\cdot\sqrt{(\Delta Q\cdot \sigma_{\Delta U})^2+ (\Delta U\cdot \sigma_{\Delta Q})^2}\cdot\left(\frac{180}{\pi}\right)
\end{eqnarray}
These analyses help visualize the changes in $P$ and $PA$ across the dust layers lying between  $(i+1)^{th}$ and $i^{th}$ star (see \citealt[][see their Figure 4 and Section 3.2 for more details on differential plots]{eswaraiah2019}).

\begin{figure}
\captionsetup[subfloat]{labelformat=empty}
\subfloat[\label{densitydp}]{\includegraphics[width=3.15in,height=2.70in]{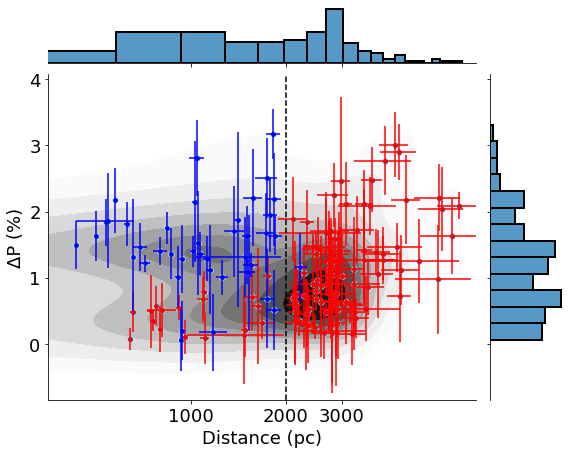}}\hfill

\subfloat[ \label{densitydpa}]{\includegraphics[width=3.15in,height=2.70in]{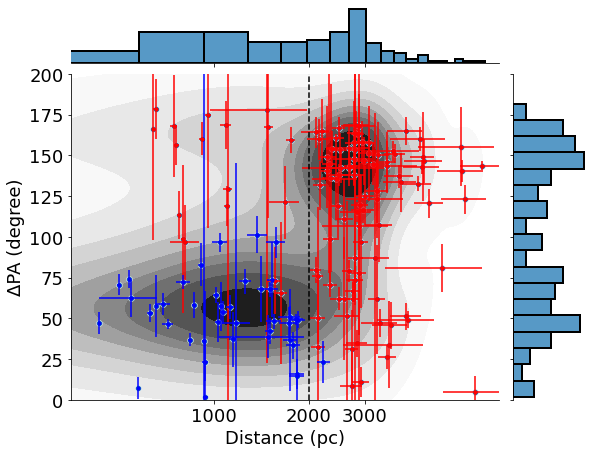}}\hfill
\caption{The differential density plots: distance {\it versus} $\Delta P$  ({\it Top}) and   $\Delta PA$  ({\it Bottom}).  The darker background is corresponding to the higher probability density, whereas the lighter density for lower probability density.  Two groups of stars: group-0 (red) and group-1 (blue) identified by the GMM are overlaid. The vertical dotted line denotes 2 kpc where the transition is observed.  Two-dimensional histograms in each panel are also shown.}
\label{differentialplots}
\end{figure}

The differential density plots  shown in the Figure \ref{differentialplots}  reveal a clear transition in both $\Delta P$ and $\Delta PA$ at a distance of $\sim$2 kpc.  To elucidate further,  the information on probability density and frequency distribution are shown with gray/black background and  histograms,  respectively. Note that while constructing histograms we use  bin sizes of 0.25\%, 10$\degr$, and 350 pc for $\Delta P$, $\Delta PA$, and  distance$_{i+1}$ respectively.

To confirm this transition occurring at a distance of $\sim$2 kpc,   we apply the Gaussian Mixture Model (GMM) analyses \citep[e.g,][]{krone2010,ducourant2017,melchior2018,cantatgaudin} on the differential plots,   by providing  $\Delta P$,  $\Delta PA$,  and,  distance$_{i+1}$ as input using the $scikit-learn$ module in {\sc python}. GMM is a probabilistic model which assumes that  the data instances are generated  from several Gaussian distributions each with an unknown set of parameters,  i.e, mean and covariance values.   Since GMM is a powerful grouping algorithm,   we use it for grouping stars based on their properties.  The GMM does this by assigning probability for each star being part of a separate group,   say group-0 or group-1.  For this purpose,  GMM uses an Expectation-Maximization (EM) algorithm.  The EM algorithm starts off by assigning the group parameters randomly to the given set of stars and is followed by two steps: expectation and maximization,  which are repeated until it achieves convergence.   In the expectation step,  EM computes the probability for each star to belong to each group based on the current group parameters.  Then in the maximization step, it updates each group based on all the stars in the group, with each star weighted by the probability that it belongs to that group.  The GMM resulting group of stars (0 and 1) are shown using red and blue symbols in Figure \ref{differentialplots}.

It is evident that two distributions exist in both plots of Figure \ref{differentialplots} and are clearly separated at $\sim$2 kpc.  The darkest probability density centred around 3 kpc could be related to the cluster members of NGC\,1893.  Note that a marked shift from a lower to higher value of $\Delta PA$ at $\sim$ 2 kpc can be seen in Figure \ref{differentialplots} ({\it Bottom}).  These findings are in accordance with the distance break obtained at 1.5$\pm$0.5 kpc based on the broken-power law and broken power-law Bayesian models (cf. Sections \ref{subsec:brokenpowerlaw} and \ref{subsec:bayesianapproach}). 

\vspace{-1em}

\subsection{Confirmation of two groups of stars}\label{3.4}
\begin{figure}
\captionsetup[subfloat]{labelformat=empty}
\subfloat[\label{propermotion}]{\includegraphics[width=3.15in,height=2.25in]{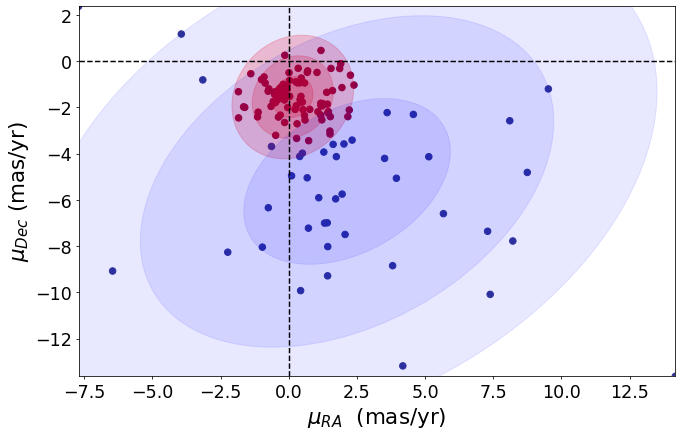}}\hfill\\
\subfloat[\label{stokesplane} ]{\includegraphics[width=3.15in,height=2.25in]{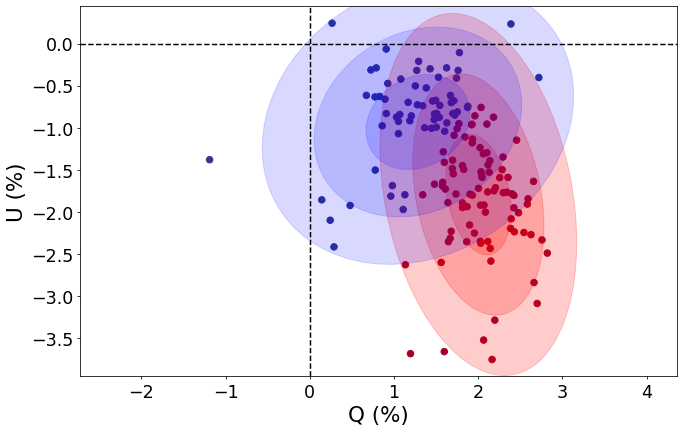}}
\caption{Proper motion ({\it Top}) and Stokes plane ({\it Bottom}) plots identified using GMM method. The  group-0 members (red) are the  cluster members and background stars, and group-1 members (blue) are the foreground stars.  The 1$\sigma$, 2$\sigma$ and  3$\sigma$ confidence ellipses of group-0 and group-1 members are also shown.}
\label{propmstokesplane}
\end{figure}

\begin{figure}
\captionsetup[subfloat]{labelformat=empty}
\subfloat[\label{propmravsdistance}]{\includegraphics[width=3.15in,height=2.25in]{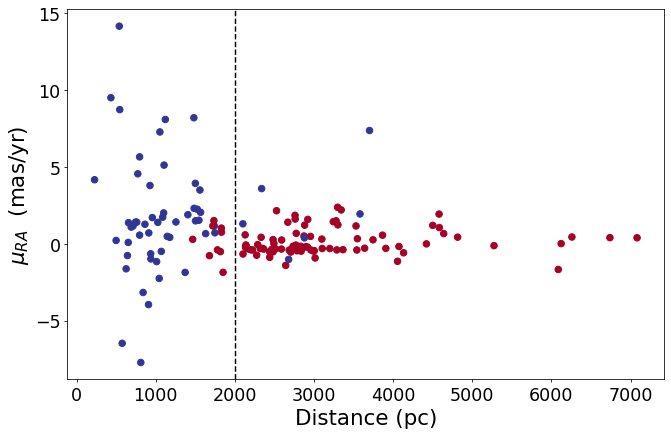}}\hfill\\
\subfloat[\label{propmdecvsdistance} ]{\includegraphics[width=3.15in,height=2.25in]{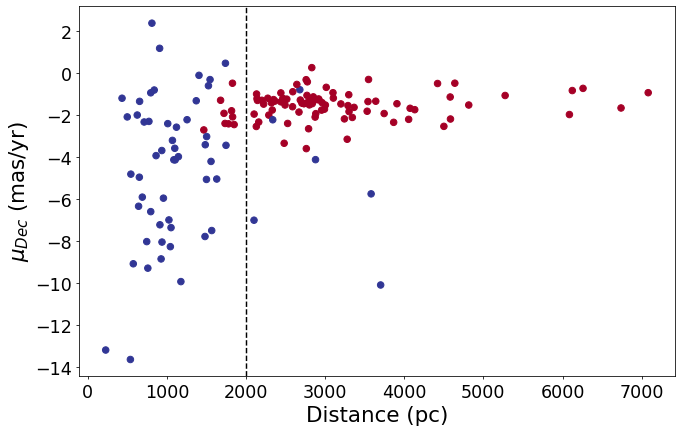}}
\caption{Distance {\it versus} $\mu_{RA}$  ({\it Top}) and  $\mu_{Dec}$ ({\it Bottom}).  In both figures (top and bottom panels),  most of the  group-0 members (red) are located after 2 kpc indicating that they are the cluster members and background stars, whereas most of the  group-1 members (blue) are located up to 2 kpc which indicates that they are the foreground stars. The black vertical dotted line denotes distance of 2kpc. }
\label{radecdistance}
\end{figure}

To confirm the presence of two groups of stars  as proposed in Section \ref{3.3},  we again employ GMM analyses. For this we utilize proper motion ($\mu_{\mathrm{RA}}$ and $\mu_{\mathrm{Dec}}$), Stokes parameters ($Q$ and $U$), and stellar parallax data as  input parameters. The proper motion data are obtained from the {\it Gaia} EDR3 catalogue, whereas the Stokes parameters using equation \ref{stokesvalues}. The GMM resulting groups (0 and 1) are shown using red and blue symbols in Figures \ref{propmstokesplane} and \ref{radecdistance}.  The 1$\sigma$, 2$\sigma$ and  3$\sigma$ confidence ellipses of group-0 and group-1 stars are also shown in Figure \ref{propmstokesplane}.

Figure \ref{propmstokesplane} ({\it Top}) shows the clear distributions predicted by GMM ~--~ cluster members and background stars as depicted in red exhibit a smaller amounts of proper motion as they fall close (0, 0), whereas the foreground stars, shown with blue, exhibit larger amounts of proper motion with a scattered distribution. This is also evident from the larger confidence ellipses for foreground stars when compared to cluster and background stars.

Similarly, the Stokes plane diagram ($Q$ vs $U$) shown in Figure \ref{propmstokesplane} ({\it Bottom})  also reveals two distributions. 
Cluster members shown in red are grouped together as their polarization properties are similar. This is because they are believed to be formed out of a same parental molecular cloud at the similar ages, and also their distances and amounts of extinction will also be similar. Owing to these, when the light from  cluster members passes through the common layers of aligned dust grains and magnetic fields, their polarization properties will be similar and as a result they exhibit clustered distribution in the Stokes plane. In case of a negligible amount of polarizing dust in the background medium (that lies between cluster members and background stars), their position in  the Stokes plane may merge with (or scatter from) those of cluster members given the magnetic field orientation is similar (or different) from that in the cluster. Foreground stars  with blue fall close to (0, 0) and also show a scattered distribution in the Stokes plane  depending upon the amount of extinction (hence polarization) present in the foreground medium and the morphology of magnetic field in the foreground medium \citep[eg.,][]{Eswaraiahetal2011}. Therefore, distance versus proper motions plots (Figure \ref{radecdistance}) demonstrate two groups being separated at 2 kpc, with a very few stars from one group falling into the another one.

\section{Results }\label{section4}
\subsection{$A_{V}$ {\it versus} $P$/$A_{V}$}\label{4.1}

Based on the broken power-law fitting on  $A_{V}$~--~$P$/$A_{V}$, shown in Figure \ref{brokenpowerlawplots} ({\it Top}), we find that the polarization efficiency ($P$/$A_{V}$) remains constant up to a threshold extinction, $A_{V}^b$, of  0.96 $\pm$ 0.09 mag with a power-law index of 0.01 $\pm$ 0.19. For $A_{V} > 0.96$ mag the dust polarization efficiency decreases with a power-law index of $-$1.0 $\pm$ 0.1. The  Bayesian analysis also reveals similar results as those obtained from broker power-law fits (see Table \ref{tab:bayesbrokenpowerlawparams}) with $A_{V}^b$ occurring at 0.9 mag with a 97 \% highest density interval (HDI) of 1 mag as shown in Figure \ref{bayesianplots} ({\it Top}). From this, we infer the existence of two dust populations. One with  $A_{V}$ < 0.9 mag exhibits a higher polarization efficiency than the other with $A_{V}$ > 0.9 mag.

\subsection{$d$ {\it versus} $P$/$d$}\label{4.2}

The broken power-law over the data of $d$~--~$P/d$, shown in Figure \ref{brokenpowerlawplots} ({\it Top}), reveals a steeply decreasing trend by the rate of polarization up to a break distance, $d^{b}$, of 1.5 $\pm$ 0.3 kpc by following an index of $-$0.8 $\pm$ 0.1. Thereafter, the rate of polarization still decreases, but with a shallower power-law index of $-$0.3 $\pm$ 0.1. The Bayesian analysis on this plot also exhibits similar results as those obtained from broken power-law fits as per the Table \ref{tab:bayesbrokenpowerlawparams}.  Therefore, we confirm the presence of two dust populations, as proposed in Section \ref{4.1}, that they are separated at the distance of $\sim$1.5 kpc.  The first dust population span below 1.5 kpc exhibits weaker rate of polarization, whereas the second dust population with distance $>$ 1.5 kpc exhibits a higher rate of polarization.

Hereafter, we consider 2 kpc as the boundary where the broken power-law and Bayesian analyses (Sections \ref{subsec:brokenpowerlaw} and \ref{subsec:bayesianapproach}), differential plots (Section \ref{3.3}), and GMM analyses on Stokes parameters and proper motion data (Section \ref{3.4}), altogether hint a change in the polarization properties and dust extinction.
Based on these, we categorize stars with $d  < 2$ kpc and $A_{V} <$ 1 mag as the foreground stars tracing the foreground dust,  whereas the stars with $d  >$ 2 kpc and $A_{V} > 1$ mag as cluster members and background stars tracing Perseus arm dust  (hereafter, Perseus dust) towards the line of sight of NGC 1893.

\subsection{Foreground contribution and intrinsic polarization measurements of the intra-cluster medium}\label{3.2.1}

Power-law indices of $A_{V}$~--~ $P/A_{V}$ (Section \ref{4.1}) imply that although foreground dust is a major contributor to polarization, it is not a sole contributor, and that there is a non-neglible contribution from Persues-dust. In contrast,  the power-law indices of $d$~--~$P/d$ (Section \ref{4.2}) suggest a higher polarization rate for Perseus-dust than foreground dust. Therefore, current observations indeed can reveal the polarization properties of intra-cluster medium of NGC\,1893 if we do the foreground correction. Based on this assumption of a small and non-negligible polarization contribution from the intra-cluster medium, we subtract the contribution of foreground polarization and extract the intrinsic polarization properties ($P_{int}$ and $PA_{int}$).

To estimate the foreground polarization measurements, we perform weighted linear fits on the Stokes parameters versus distance plots by considering the data up to 2 kpc. From the fit we have estimated the resultant Stokes parameters at 2 kpc, which are treated as foreground Stokes parameters, $Q_{\mathrm{fg}}$ and $U_{\mathrm{fg}}$. These are vectorially subtracted from the Stokes parameters of stars distributed beyond  2 kpc and obtain the intrinsic Stokes parameters using the following relations:

\begin{eqnarray}\label{qintuint}
                     Q_{int} = Q - Q_{fg}\nonumber\\
                     U_{int} = U - U_{fg}
\end{eqnarray}

\begin{figure}
\includegraphics[width=3.15in,height=3in]{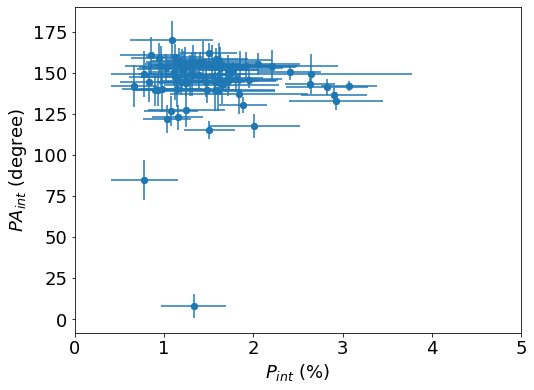}
\caption{$P_{int}$ {\it versus} $PA_{int}$  scatter plot after applying signal-to-noise ratio condition with the criteria of $P_{\mathrm{int}}$/$\sigma_{P_{\mathrm{int}}}$~$<$~2 }
\label{scatterwoerror}
\end{figure}

The intrinsic polarization ($P_{int}$) and intrinsic polarization angle ($PA_{int}$)  of the Perseus-dust and its corresponding errors are derived 
using equations \ref{eq14} and \ref{eq15}.  We have excluded 21 stars with lower signal-to-noise ratio with criteria of $P_{\mathrm{int}}$/$\sigma_{P_{\mathrm{int}}}$~$<$~2. The resulting $P_{int}$ and $PA_{int}$ values are shown in Figure \ref{scatterwoerror}.
To infer the variation in the intrinsic polarization properties of the intra-cluster medium, we further exclude 38 stars whose distances lie between 2 and 3 kpc. We assume that the remaining 32 stars lying beyond 3 kpc can trace the dust properties of the intra-cluster medium.

\subsection{Effect of foreground polarization and magnetic field morphology in  NGC\,1893}\label{4.3}

\begin{figure}
\captionsetup[subfloat]{labelformat=empty}
\subfloat[\label{pdfcdfp}]{\includegraphics[width=8cm,height=6cm]{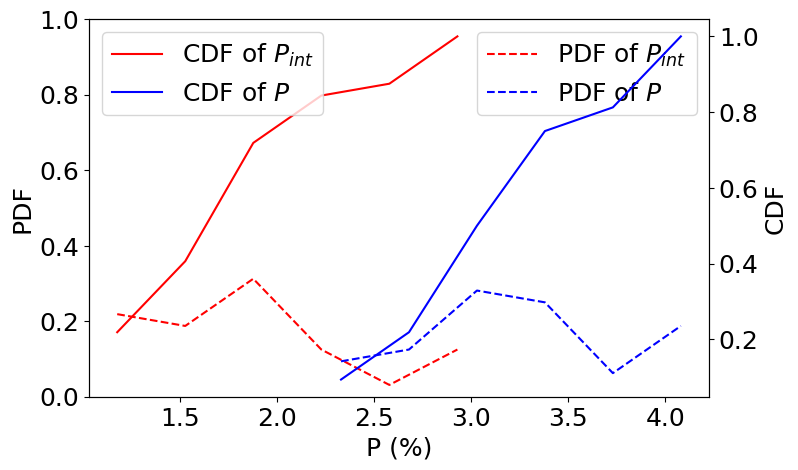}}\hfill\\
\subfloat[\label{pdfcdfpa}]{\includegraphics[width=8cm,height=6cm]{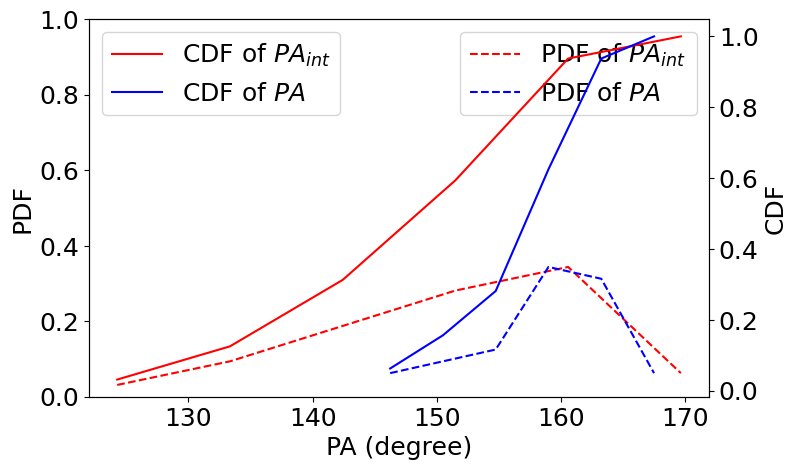}}
\caption{Probability Density Function (PDF) and Cumulative Distribution Function (CDF) of $P$ , $P_{int}$ ({\it Top}) and  $PA$ , $PA_{int}$ ({\it Bottom}) respectively.}
\label{ppapdfddf}
\end{figure}

To examine the effect of foreground polarization subtraction, we construct cumulative distribution function (CDF) and Probability distribution function (PDF) for the polarization measurements before ($P$ and $PA$) and after ($P_{int}$ and $PA_{int}$) foreground subtraction.  These CDF and PDF plots for $P$ and $P_{int}$ and $PA$ and $PA_{int}$ are shown in Figure \ref{ppapdfddf}. 
To verify whether the mentioned pair of distributions ($P$ {\it versus} $P_{int}$ and $PA$ {\it versus} $PA_{int}$) are similar or not, we perform Kolmogorov-Smirnov (KS) test on the PDFs and CDFs.  KS test yield two parameters~--~statistic and p-value, of which p-value is the most important parameter as it infers whether a pair of  distributions are drawn from a same parental distribution or not.  A smaller p-value denotes that the two distributions are different, such that they are not drawn from a same parental distribution.  KS-test on CDFs and PDFs of $P$ vs. $P_{int}$ and $PA$ vs. $PA_{int}$  suggest smaller p-values (6.8e-10 and 5.2e-05).  These imply  $P$ and $PA$ are entirely different from $P_{int}$ and $PA_{int}$, respectively, and are not drawn from the same distributions. 

\begin{figure}
\captionsetup[subfloat]{labelformat=empty}
\subfloat[\label{histgaussianp}]{\includegraphics[width=9cm,height=6cm]{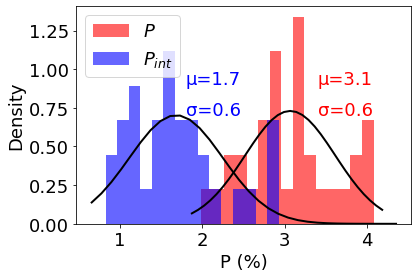}}\hfill\\
\subfloat[\label{histgaussianpa}]{\includegraphics[width=9cm,height=6cm]{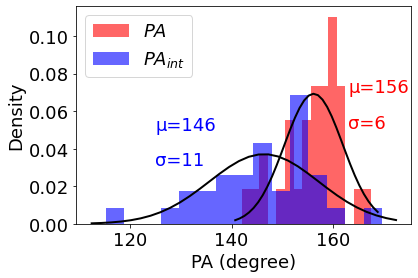}}
\caption{Histogram with Gaussian fits(shown in figure as a thick black line) of $P$ , $P_{int}$ ({\it Top}) and $PA$ , $PA_{int}$ ({\it Bottom}). The mean values and the standard deviations corresponding to $P$ ,$P_{int}$  and $PA$, $PA_{int}$ are shown in red and blue colour fonts respectively.}
\label{histgaussianppa}
\end{figure}

The above findings are further corroborated by the observed different Gaussian means for polarization and polarization angles, as indicated in Figure \ref{histgaussianppa}. We observe two noticeable differences in the intrinsic polarization properties of the intra-cluster medium after the foreground subtraction: (a) the Gaussian mean of $P_{int}$ is systematically smaller than that of $P$, as clear from Figure \ref{histgaussianppa} ({\it Top}) due to  removal of the foreground contribution, and (b) Gaussian means of $PA_{int}$ and $PA$ are similar within the standard deviations albeit with a wider $PA_{int}$ distribution indicating slight change in the magnetic field morphology in NGC\,1893. 

\begin{figure*}
\centering
\captionsetup[subfloat]{labelformat=empty}
\begin{minipage}{1\textwidth}
\centering
\hspace*{\stretch{1}}
\subfloat[]{\includegraphics[width=0.49\textwidth]{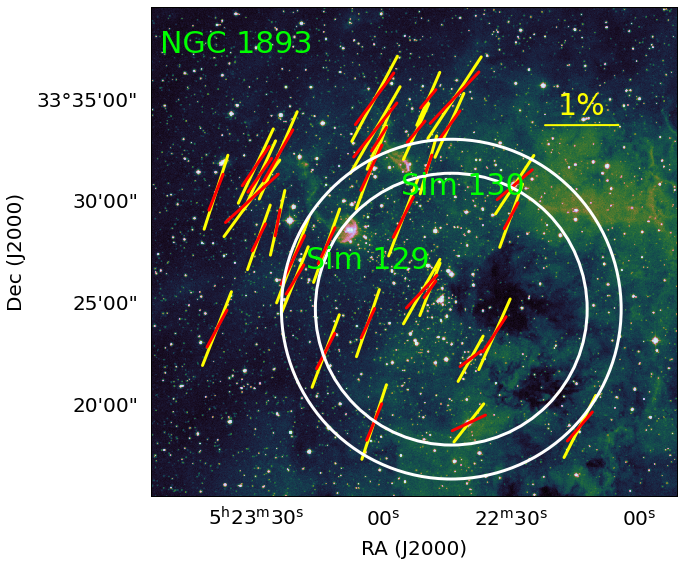}}%
\hspace{\stretch{2}}
\subfloat[]{\includegraphics[width=0.49\textwidth]{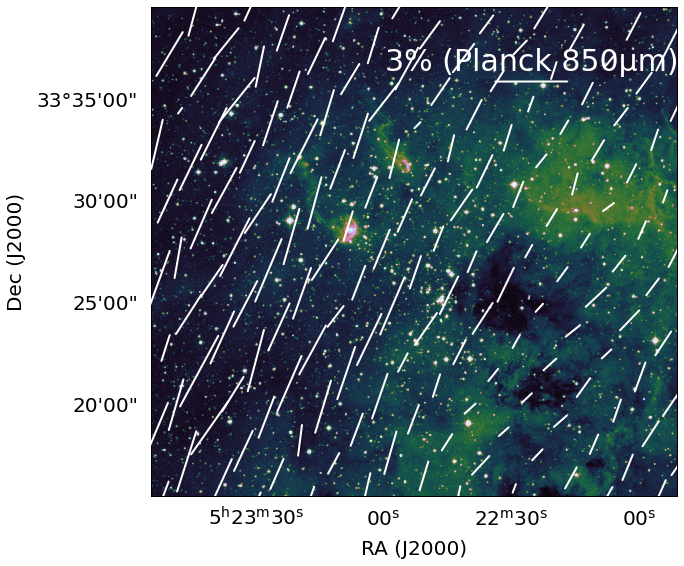}}%
\hspace{\stretch{1}}
\caption{{\it Left}: Intrinsic magnetic field map of NGC 1893. The red coloured lines are the polarization vectors before the foreground subtraction, and the yellow coloured lines are the polarization vectors after the  foreground subtraction. The length of the vector is proportional to the degree of polarization ($P$ or $P_{int}$), whereas the orientation of the vector corresponds to the magnetic field orientation ($PA$ or $PA_{int}$) with reference to equatorial north but increasing towards the east. Two concentric circles of radii 400\arcsec and 500\arcsec are shown. The region spanning between the two concentric circles denotes the $\sim$400$\arcsec$~--~500$\arcsec$ region. The location of two emission nebulae Sim 129 and Sim 130 are also shown.   {\it Right}: Oversampled (pixel size 1$^\prime$) magnetic field map morphology inferred based on {\it Planck}/353 GHz dust polarization data. The {\it Planck} magnetic field vectors (white) can be seen to match with those inferred from $I$-band polarization data presented in the left panel.}
\label{magmap}
\end{minipage}
\end{figure*}

Figure \ref{magmap} ({\it Left}) shows the polarization vector map delineating the magnetic field morphology before (red) and after (yellow) foreground subtraction. The length of the vector is proportional to the degree of polarization ($P$ or $P_{int}$), whereas the orientation of the vector corresponds to the magnetic field orientation ($PA$ or $PA_{int}$) with reference to equatorial north but increasing towards the east.
Therefore, the effect of foreground subtraction on the polarization properties of intra-cluster medium can be seen as a systematic reduction in fraction of polarization but without a significant difference in the overall mean magnetic field orientation in NGC\,1893.  Moreover,  magnetic field in the foreground (circular mean $=$ 156$\degr$ $\pm$ 6$\degr$) and intra cluster medium (circular mean $=$ 147$\degr$ $\pm$ 11$\degr$) are nearly similar within the uncertainties, indicating the presence of uniform magnetic field orientation towards the direction of NGC\,1893.  The overall mean field orientation in the intra-cluster medium is nearly parallel to the position angle of the Galactic plane, $141\degr$, at $b=-1.53\degr$ towards the direction of NGC\,1893. The mean magnetic field orientation inferred by {\it Planck} dust continuum polarization data at 850 $\mu$m \footnote{The {\it Planck} polarization data comprising of Stokes $I$, $Q$, and $U$ maps were  extracted from {\it Planck} Public Data Release 2 \citep[][]{planck2016b} of Multiple Frequency Cutout Visualization (PR2 Full Mission Map with PCCS2 Catalog; {\url{https://irsa.ipac.caltech.edu/applications/planck/}}. The  pixel size and beam size were  $\sim$1$\arcmin$ and $\sim$5 $\arcmin$, respectively. We have processed the data and obtained the magnetic field map using the procedures presented in Section 3.4 of \cite{Baug2020} and references therein.}  is found to be 153 $\degr$ $\pm$ 9 $\degr$ \citep[][]{planck2015}. The overall magnetic field morphology, based on {\it Planck}/353 GHz low-resolution (5\arcmin) dust polarization data, is found to be uniform as shown in the Figure \ref{magmap} ({\it Right}). This overall geometry is consistent with the magnetic field morphology inferred from optical data. 

\subsection{Changes in the intrinsic polarization parameters and other relevant parameters in the cluster region}\label{4.4}

\begin{figure*}
    \centering
    \includegraphics[width=10.5cm,height=17cm]{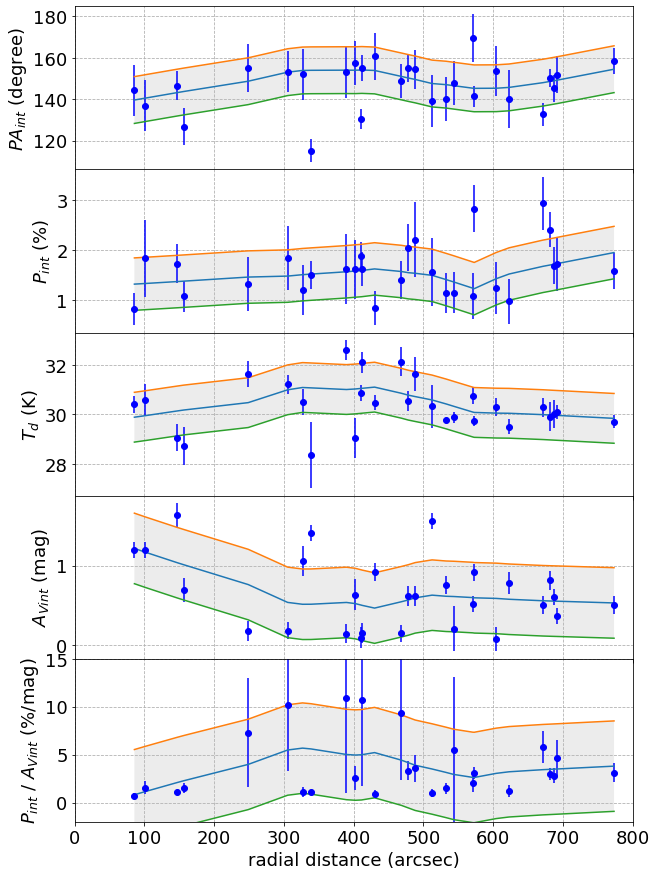}
    \caption{Variation of radial distance {\it versus}  (a) $PA_{int}$, (b) $P_{int}$, (c) $T_{d}$, (d) $A_{Vint}$,  and (e) $P_{int}/A_{Vint}$ for the cluster members.  The panels  (a)~--~(d) consist of 28 stars, whereas panel (e) has 26 stars.  Four stars which do not have reliable $A_{V}$ values are removed from all the panels,  and two additional stars with large errors in $P_{int}/A_{Vint}$ are  removed from panel (e).  The overall trends traced by the LOWESS-smoothed curves (blue) are shown in all the panels. The corresponding 1-$\sigma$ regions (grey) bounded by the +$\sigma$ (orange) and -$\sigma$ (green) lines are also shown.
    }
    \label{radplot}
\end{figure*}

In order to examine the change in the polarization parameters in response to other crucial parameters in the star forming region, we have plotted radial distance (with respect to central coordinates of NGC\,1893) {\it  versus} $PA_{int}$, $P_{int}$, $T_{d}$, $A_{V_{int}}$, and $P_{int}/A_{V_{int}}$ as shown in Figure \ref{radplot}. Here, $A_{V_{int}}$ is the intrinsic total extinction of stars with distance greater than 3 kpc after correcting for a mean foreground extinction of 0.97 mag. $T_{d}$ is the dust temperature and $P_{int}/A_{V_{int}}$ is the intrinsic polarization efficiency.   To make the trends in the plot more clear, we over plot  Figure \ref{radplot} with curves drawn using the locally weighted scattered smoothing (LOWESS) algorithm \citep[][]{Cleveland1979}{} from $statsmodels$  package in {\sc python}. We use a `frac' parameter of 0.5 to draw the LOWESS curves over all the panels. The corresponding 1-$\sigma$  regions (green) bounded by the $+$$\sigma$ (orange) and $-$$\sigma$ (green) lines are also shown in all the plots.

From Figure \ref{radplot}, the magnetic field orientation, based on the majority values of $PA_{int}$, is found to be rather uniform without a significant change in the intra cluster medium of NGC\,1893 region.  The $P_{int}$ shows a constant trend up to $\sim$500$\arcsec$, and afterwards it exhibits a scattered distribution. Dust temperature is tending to increase up to $\sim$500$\arcsec$, and thereafter it decreases. $A_{V_{int}}$ shows a decreasing trend up to $\sim$400$\arcsec$, and thereafter it remains nearly constant. The intrinsic polarization efficiency, $P_{int}/A_{V_{int}}$, exhibits an increasing trend  up to $\sim$400$\arcsec$, and thereafter it shows a scattered distribution.  Except for  $PA_{int}$ and $P_{int}$,   all parameters show a clear change within the region spread over $\sim$400$\arcsec$~--~500$\arcsec$ in NGC\,1893. This is  where the two nebulae Sim 129 and Sim 130 are forming due to radiative driven implosion by the UV radiation emanating from the O/B- type stars \citep[][]{ Maheswaretal2007,sourabhsharma2007}.  Changes in various parameters are found to be happening in the area lying between two concentric circles of radii 400\arcsec and 500\arcsec as shown  in Figure  \ref{magmap} ({\it Left}). 
We note here that both tips of Sim 129 and 130, and the shell-like structure surrounding the cluster centre, seem to be formed as a consequence of swept material out of expanding ionization and shock fronts from the O/B type stars formed near the cluster centre.

\section{Discussion}\label{section5}

From the broken power-law fit and Bayesian analyses of  $A_{V}$~--~$P$/$A_{V}$ (see Figures \ref{brokenpowerlawplots} and \ref{bayesianplots} and Section \ref{4.1}) we show that polarization efficiency ($P$/$A_{V}$) exhibits a constant trend as a function of extinction ($A_{V}$) up to threshold extinction value $A_{V}^b$~$\sim$0.9 mag with a power-law index of $\sim$0  (see, Table \ref{tab:bayesbrokenpowerlawparams}).   This constant trend implies all the dust grains are aligned uniformly, and they polarize the stellar light up to the same degree at all the dust layers located up to $A_{V}$$\sim$0.9 mag .
For $A_{V}$ $>$ 0.9 mag, polarization efficiency declines steeply with 94\% highest density interval of indexes ranging from $-1.2$ to $-0.7$ (see Figure \ref{bayesianpdf} ({\it Top})),  indicating poorly aligned dust grains.  A number of  studies  have probed the variation of $P/A_{V}$ as a function of $A_{V}$ towards several dark clouds  like Taurus and Rho Ophiuchus  molecular cloud complexes using near-infrared observations.   Majority of these studies find a drop in $P/A_{V}$ as a function of  $A_{V}$ with a typical power law index close to  $-$0.5 \citep[eg.,][]{tamura1987,gerakines1995,whittet2008,chapman2011,cashman2014}.   Some studies \citep[eg.,][]{goodman1995} find a constant $P$ {\it versus} $A_{V}$ trend, which in terms of  $P/A_{V}$ {\it versus} $A_{V}$ will follow a steeply decreasing trend, indicating poor dust grain alignment at relatively higher opacities.   \citet{arceetal1998} have found a linear and flat trends in their $P$ {\it versus} $A_{V}$ for the stars lying behind the diffuse and dense regions of Taurus, respectively. However, their data in terms of $P/A_{V}$ {\it versus} $A_{V}$, will follow similar trends that we observed in this work. Similarly, \citet{jones2014} while studying the variation of fractional polarization per unit optical depth ($P/\tau$) {\it versus} $A_{V}$ towards a selection of starless cores using NIR and submm observations also  find a change in the power-law from $-$0.5 to $-$1 at $A_{V}$~$\sim$20 mag which is again similar to our trends.   Another study carried out by \citet{WangJWetal2017} showed that broken power-law prevails towards IC\,5146 molecular cloud and evidenced a better dust grain alignment at relatively denser regions of the cloud with $A_{V}~>~ 2.25$ mag in comparison to the poorly aligned dust grains in the low-density region with $A_{V} <$ 2.25 mag. This result is in contrast to our result of efficiently aligned dust grains in the diffuse ISM and poorly aligned Perseus-arm dust.

\begin{figure}
    \centering
    \includegraphics[width=9cm,height=7cm]{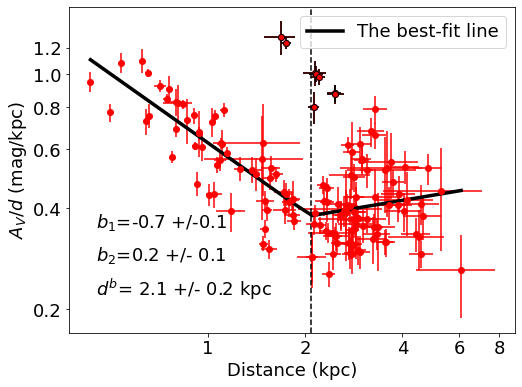}
    \caption{ Distance {\it versus} $A_{V}/d$ plot.  The Best fit broken power-law is denoted with a thick line. The dotted vertical line corresponds to  $d^{b}$ at which the power-law breaks. Eight scattered points shown with black error bars are excluded while performing the fit. The Best-fit parameters are also overlaid. }
    \label{avbydbroken}
\end{figure}

\begin{figure}
    \centering
    \includegraphics[width=9cm,height=7cm]{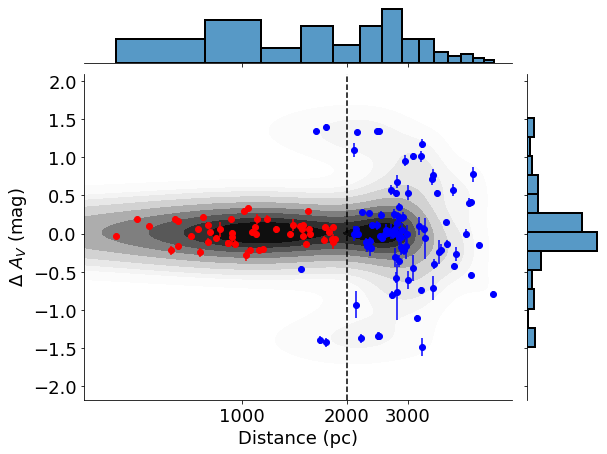}
    \caption{ The differential density plot: distance {\it versus} $\Delta A_{V}$.  The darker background is corresponding to the higher probability density, whereas the lighter density for lower probability density.  Two groups of stars: foreground stars (blue) and cluster members and background stars (red) identified by the GMM are overlaid. The vertical dotted line denotes 2 kpc where the transition is observed.  The  constant and scattered distributions of $\Delta A_{V}$  corresponds to the uniform and non-uniform dust distributions before and after 2 kpc, respectively.   Two-dimensional histograms in each panel are also shown.
    }
    \label{deltaav}
\end{figure}

From the broken power-law fit and Bayesian analysis of $d$~--~$P$/$d$ as presented in section  \ref{4.2}, it is seen that the rate of polarization  decreases steeply as a function of distance up to a threshold distance of $\sim$2 kpc with a relation $P/d$~$\propto$~$d^{-0.8}$, thereafter $P/d$ decreases but with a shallower rate with $P/d$~$\propto$~$d^{-0.3}$. We found that similar power-law indices have been found by the data on distance versus rate of extinction,  $d$ {\it versus} $A_{V}/d$ plot as shown in Figure \ref{avbydbroken}.  The data up to $\sim$2 kpc follow the relation $A_{V}/d$~$\propto$~$d^{-0.7}$, while beyond $\sim$2 kpc they exhibit $A_{V}/d$~$\propto$~$d^{0.2}$. Therefore, the observed trends in the rate of extinction could be reflecting on the rate of polarization because both the quantities, extinction and polarization, are related.   However, there exist a difference in the indices of the relations $P/d \propto d^{-0.3}$ and $A_{V}/d \propto d^{0.2}$ for stars with > 2 kpc. This implies that although the dust extinction increases abruptly beyond 2 kpc, the amount of polarization does not increase in the same proportion. We suggest that the rate of polarization plot reveals whether the dust grains are distributed in a uniform or non-uniform manner as a function of distance. To confirm this, we have plotted distance$_{i+1}$ {\it versus} differential extinction ($\Delta A_{V}$) in Figure \ref{deltaav}, which essentially reveals a constant and scattered distribution of $\Delta A_{V}$  corresponding to uniform and non-uniform dust distributions before and after 2 kpc, respectively.

Based on the analyses $A_{V}$~--~$P$/$A_{V}$, $d$~--~$P$/$d$, and the differential plots,  we propose that there exist two dust populations: (i) the foreground dust distributed between local Solar neighbourhood and Perseus arm and is ~--~ located up to $\sim$2 kpc and characterized by $A_{V}$ < $\sim$1 mag,   and (ii) the Perseus arm dust (Perseus-dust) could be part of the Perseus spiral arm in the direction of NGC\,1893 and is ~--~ distributed beyond $\sim$2 kpc and characterized by $A_{V}$~>$\sim$1 mag.  Foreground dust is found to be uniformly distributed and efficiently aligned to the same degree at all distances to polarize the stellar light efficiently, as clear from the flat distribution in $P/A_{V}$ versus $A_{V}$ plot. In contrast, Persues-dust seems to be non-uniform and not efficient enough to produce the polarization, as clear from the steeply decreasing trend in $P/A_{V}$ {\it versus} $A_{V}$. Therefore, rate of polarization plot reveals about the distribution of dust, whereas the polarization efficiency plot sheds a light on dust grain alignment efficiency.  We suggest that the observed differences in the polarization efficiency of the foreground and Perseus-dust are not due to the complex magnetic fields (but because of the differences in the dust distribution and their respective polarization efficiencies) as the magnetic field is found to be uniform across the entire line of sight to NGC\,1893.

Although Perseus-dust seems to be non-uniform and poorly aligned, we assume that it infers the intra-cluster medium properties with a proper treatment of foreground subtraction and excluding stars with distance less than 3 kpc as well as lower SN ratio of intrinsic polarization (see Section \ref{4.3}). 
From Figure \ref{radplot}, we notice that  $T_{d}$, and $P_{int}/A_{V_{int}}$ parameters tend to increase up to $\sim$400$\arcsec$~--~500$\arcsec$ distance from the cluster centre, but the total dust extinction, $A_{V_{int}}$, decreases up to $\sim$400$\arcsec$~--~500$\arcsec$ distance.   $P_{int}$ at the same time tends to follow a constant trend up to $\sim$500$\arcsec$.~But interestingly, the magnetic field orientation in the cluster region, as inferred by $PA_{int}$, is found to be nearly constant at all the radial distances. This implies that the increasing trend seen in $P_{int}/A_{V_{int}}$ is not because of spatial variation in $A_{V_{int}}$ and of a constant trend in $PA_{int}$.  Instead, it is most likely caused by an increase in the high energy stellar radiation emanating from the O/B type stars at the cluster center. Two emission nebulae Sim 129 and 130 are formed by the photoevaporation of the pre-existing clouds as a consequence  of advancing ionization and shock fronts along with high energy UV photons and stellar winds originating from the O/B-type stars at the center of NGC\,1893 cluster  \citep[see,][]{Maheswaretal2007,sourabhsharma2007}.   It is also possible that a number of newly formed pre-main-sequence stars around these nebulae could also act as secondary sources of radiation in the vicinity of Sim 129 and 130 so as to alter the dust grain alignment.

Such an increase in the amount of radiation towards the two nebulae can also increase the dust temperature $T_{d}$. This would  result in a better alignment of the dust grains in accordance with the RAT mechanism \citep{tram2021}. Consequently, this in turn would be resulting in a higher polarization efficiency ($P_{int}/A_{V_{int}}$) towards the two nebulae. 
Therefore, increase in the amount of radiation could be the plausible reason behind the observed increase in the two parameters  $T_{d}$, and $P_{int}/A_{V_{int}}$ close to the two emission nebulae in the region distributed between  $\sim$400$\arcsec$~--~500$\arcsec$ as represented with two concentric circles in Figure \ref{magmap} ({\it Left}).

\begin{figure}
    \includegraphics[width=9cm,height=7cm]{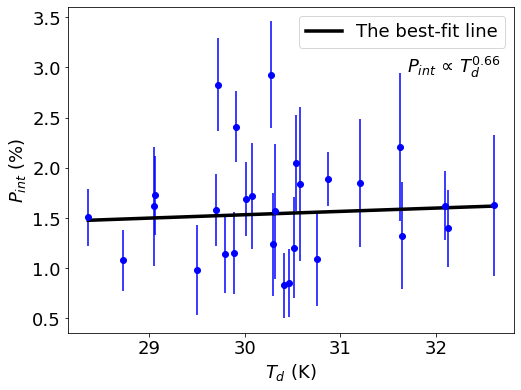}
    \caption{ $T_{d}$ {\it versus} $P_{int}$ plot with the best fit power-law (thick line). The power-law relation is also overlaid.
    }
    \label{pintvstint}
\end{figure}

We further performed a power-law fit to the $T_{d}$ {\it versus} $P_{int}$ data to quantify the existence of any positive correlation.  As shown in Figure \ref{pintvstint}, the data exhibits a positive power-law index of 0.66 indicating slow increase in $P_{int}$ as a function of increase in $T_{d}$.   To confirm this correlation,  we perform the spearman rank-order correlation test using the $scipy-stats$ module in {\sc Python}. The spearman test gives two outputs: correlation coefficient (r)  and probability for an uncorrelated system of datasets (p-value), which are given to be 0.139 and 0.48, respectively. The small positive r-value and statistically insignificant p-value (as p > 0.05) indicates a very weak correlation between $T_{d}$ and $P_{int}$.

Nevertheless, within a range,   $\sim$29.5~--~31 K,   of  $T_{d}$,    the $P_{int}$ is seen to  exhibit a scattered distribution. Within this range of temperatures, some dust grains show  relatively higher $P_{int}$ values  ($P_{int}$ > 2.3 \%) probably due to better alignment by RAT mechanism.

To confirm this, we have examined offset angle ($\psi$) between the magnetic field and the radiation field emanating from two O/B-type stars (HD 242935 and BD +33$\degr$ 1025 that are located at the center of the NGC 1893 cluster) for three stars whose $P_{int}$ values are higher than 2.3\%.  The $\psi$ angles are found to be 63$\degr$,  80$\degr$,  and 68$\degr$ and  in the order of increasing level of $P_{int}$. According to \citet{Lazarian2021}, for $\psi$>45$\degr$ the fraction of dust grains driven to high J-attaractor points increase significantly due to weaker rotational disruption and relatively strong RAT, thereby resulting in a higher fraction of polarization ($P_{int}$).  Therefore, three stars with $\psi > 45 \degr$ exhibit relatively higher $P_{int}$ could be due to enhanced RAT mechanism.

\section{summary and conclusions}\label{section6}

The main aim of this paper is to investigate whether polarization observations from a distant target enable us to reveal the dust properties and magnetic field orientation within the target of interest. For this purpose, we have studied the variation of interstellar polarization as a function of extinction and distance towards the direction of NGC 1893. We have utilized  the newly acquired $I$-band polarization data from ARIES IMaging POLarimeter (AIMPOL) plus the already published $I$-band polarization data \citep{Eswaraiahetal2011}.  In addition, we also utilize distances from \citet{Bailer-JonesRybizkiFouesneauetal2021} and extinction data 
from \citet{Greenetal2019}.  We have analysed two key plots (i) polarization efficiency ($P/A_{V}$) as a function of extinction ($A_{V}$) and (ii) rate of polarization ($P/d$) as a function of distance ($d$) by employing fitting  (i) broken power-law and (ii) Bayesian analysis on both of the plots. 

The following are the main conclusions of our study:

\begin{itemize}

\item The analyses on extinction ($A_{V}$) {\it versus} polarization efficiency ($P/A_{V}$) plot reveals a flat trend with a power-law index of $\sim$ 0 up to the threshold extinction, $A_{V}^{b}$, of 1 mag.
For $A_{V}$~$>$~$\sim$1 mag, $P/A_{V}$ exhibits a steeply decreasing trend with a power-law index ranging from $-$1.2 to $-$0.7.  \\

\item The analyses on distance ($d$) {\it versus} rate of polarization ($P/d$) suggest  that $P/d$ declines steeply with an index of $-$0.8 up to the threshold distance, $d^{b}$, of $\sim$1.5 kpc and beyond which the index is found to be $-$0.6 indicating same declining rate of polarization but with a lesser degree. \\

\item Based on the analyses of $A_{V}$~--~$P/A_{V}$ and $d$~--~$P/d$,  we suggest that there exist two dust populations: (1) foreground dust characterized by $A_{V}$~$<$~$\sim$1 mag and distributed up to $\sim$2 kpc distance, and (2) Perseus-arm dust (Perseus dust) characterized by $A_{V}$~$>$~$\sim$1 mag and distributed beyond $\sim$2 kpc. \\

\item The relations $A_{V}$~--~$P/A_{V}$ and $d$~--~$P/d$ reveal two different clues about the dust grain properties: the former is about dust grain alignment efficiency and the latter is on the distribution of dust.\\

\item We find that foreground dust is uniformly distributed and  efficiently aligned, whereas the Perseus-dust is non-uniformly distributed and poorly aligned.\\

\item After correcting for foreground polarization, we have derived the intrinsic polarization parameters ($P_{int}$ and $PA_{int}$) and other parameters such as intrinsic extinction ($A_{V_{int}}$), dust temperature ($T_{d}$), and intrinsic polarization efficiency ($P_{int}/A_{V_{int}}$). \\

\item The magnetic field orientation within the star forming region is found to be almost constant, and is not too different from that in the foreground medium. \\

\item  We notice an increasing trend in  $T_{d}$, and $P_{int}/A_{Vint}$ towards the two nebulae Sim 129 and Sim 130, which we attribute to the increase in the stellar radiation emanating from the O/B-type stars located near the cluster centre of NGC\ 1893 or to the  pre-main-sequence stars in the vicinity of the two nebulae which can act as secondary sources of radiation.\\

\item $T_{d}$ and $P_{int}$ shows a very weak correlation,  however a fewer data points show relatively higher polarization could be due to RAT mechanism. \\
\end{itemize}

In order to strengthen our statements, we plan to perform similar analyses using the polarization, extinction, and distance data towards other stellar clusters that are situated towards the direction of, but lying close to, NGC\,1893.

\section*{Acknowledgements}

We thank the referee for constructive suggestions which have improved the content and flow of this paper.  C.E. acknowledges the financial support from grant RJF/2020/000071 as a part of Ramanujan Fellowship awarded by Science and Engineering Research Board (SERB), Department of Science and Technology (DST), Govt. of India. C.E. and B.N thank Prem Prakash and Belinda Damian for their  help in programming and analyses. 
This work is done as part of the MS thesis project at Indian Institute of Science Education and Research (IISER) Tirupati.

\section*{Data Availability}
The partial data is given in Table \ref{sampletable} and the complete dataset can be accessed online in the Harvard Dataverse using the following  link: \url{{https://dataverse.harvard.edu/privateurl.xhtml?token=76d51b62-b334-4289-aa3c-34a265634266}}.

\bibliographystyle{mnras}

\bibliography{reference.bib} 

\appendix

\section{Bayesian analyses}\label{appendixa}
Bayesian analysis is a method of statistical inference, where the available knowledge about parameters in a statistical model is updated with the information in the observed data. The background knowledge is expressed as a prior distribution and combined with observational data in the form of a likelihood function to determine the posterior distribution \citep[][]{vandeschoot2021}. The most general form of Bayesian inference is:

\begin{eqnarray}
P(\theta/D)&=&\frac{P(\theta)\cdot P(D/\theta)}{P(D)}\\
Posterior&=&\frac{Prior\cdot Likelihood}{Evidence}
\end{eqnarray}

where $D$ is the observed data and  $\theta$ represents the model parameters.  The posterior $P(\theta/D)$ describes the probability of the model parameters matching the given data. The evidence $P(D)$ is the probability of obtaining the data, which mainly serves as a normalization factor for the posterior. The prior $P(\theta)$ serves as the initial guessing probability of the model parameters based on our prior knowledge.  The likelihood $P(D/\theta)$  describes how likely it is for a given model parameter set to match the observed data.

We already know from previous studies that the probability density function (PDF) of fractional polarization ($P$) follows a Rician distribution \citep{Riceetal1945,Wardle&Kronbergetal1974,Simmons&Stewartetal1985,Quinnetal2012}. Similarly, to determine the exact PDF distributions followed by $P/A_{V}$ and $P/d$ we use the $distfit$ package in {\sc python}. It compares the  distributions of  $P/A_{V}$ and $P/d$  with 26 continuos distributions available in $scipy-stats$,  a module in python, to find the distribution which can fit the $P/A_{V}$ and $P/d$  data best.  $distfit$  ranks the 26 candidate distributions from best to worst based on their  Residual Sum of Squares (RSS) values. As a result, the best fit distributions with the lowest values of RSS  are selected. Hence, we find that  $P/A_{V}$ follow a Student's T distribution, and has PDF of the form:
\begin{eqnarray}\label{equpbyav}
f(x/x_{0},\sigma,\nu)&=&\frac{\Gamma(\nu+1/2)}{\Gamma(\nu/2)}~ \left(\frac{\sigma}{\pi\nu}\right)^{1/2}~ \left(1+\frac{\sigma (x-x_{0})^2}{\nu}\right)^\frac{-(\nu+1)}{2}
\end{eqnarray}

where $x$ and $x_{0}$ are the observed and real values of $P/A_{V}$,  and $\sigma$ is the uncertainty in any observed value of $P/A_{V}$.  $\Gamma$ is the gamma function,   and $\nu$ is the degrees of freedom of the Student's T distribution.  Similarly, $P/d$ is found to follow a Rician distribution with the PDF given by:
\begin{eqnarray}
g(x/x_{0},\sigma)&=&\frac{x}{\sigma^2}\cdot exp\left( \frac{-(x^2 + x_{0}^2)}{2\sigma^2}\right)~I_{0}\left(\frac{x\,x_{0}}{\sigma^2}\right),
\end{eqnarray}

where $x$ and  $x_{0}$ are the observed and real values of $P/d$ and $\sigma$ is the uncertainty in any observed value of $P/d$. $I_{0}$ is the zeroth order modified Bessel function.  The likelihood functions $L$ and $L^\prime$, respectively for $P/A_{V}$ and $P/d$, can be defined as:
\begin{eqnarray}
L(x_{0},\sigma,\nu)&=&\prod_{i=1}^{\infty}f(x_{n}/x_{0},\sigma,\nu)\label{eqpbyav}\\
L^\prime(x_{0},\sigma)&=&\prod_{i=1}^{\infty}g(x_{n}/x_{0},\sigma)\label{eqpbyd}
\end{eqnarray} 
Here $x_{n}$  represents the $n^{th}$ measurement of $P/A_{V}$ in equation \ref{eqpbyav} and $P/d$ in equation \ref{eqpbyd} respectively.
To perform the fit to the two plots using Bayesian method, we assume the following broken power-law model for $A_{V}$~--~$P/A_{V}$:
\begin{eqnarray}\label{eq:bpowerlawbaye1}
x_{0}&=&\begin{cases}
 \alpha\cdot \left(\frac{z}{{A_{V}}^b}\right)^{\beta_{1}}, & \text{for}~~ z \leq {A_{V}}^b \\[10pt]
 \alpha\cdot \left(\frac{z}{{A_{V}}^b}\right)^{\beta_{2}}, &
 \text{for}~~ z > {A_{V}}^b 
 \end{cases}
\end{eqnarray}
Here $\alpha$, $\beta_{1}$, and $\beta_{2}$ are the free model parameters.  $x_{0}$ is the real value of $P/A_{V}$,  $z$ is $A_{V}$,  and $A_{V}^b$ is the $A_{V}$ at which break occurs in %$P$~--~$A_{V}$ and 
$A_{V}$~--~$P/A_{V}$. Similarly, for $d$~--~$P/d$,  we assume  a broken power-law similar to equation \ref{eq:bpowerlawbaye1}, where $x_{0}$, $A_{V}$,  and $A_{V}^b$ are replaced with $P/d$,  $d$ and,  $d^b$ i.e, the $d$ at which the  break occurs in $d$~--~$P/d$ respectively.

For each relation, we define uniform priors within reasonable limits based on the best-fit parameters and their uncertainties from broken power laws (see Section \ref{subsec:brokenpowerlaw} and Table \ref{tab:brokenpowerlawparams}). 

The following are the priors: 

(a) For $A_{V}$~--~$P/A_{V}$:
\begin{eqnarray}
P(\alpha)&=&\begin{cases}
              \text{uniform}\quad \text{if $0<\beta<5$ }\\
              $0$ \quad\text{otherwise}
              \end{cases}\nonumber\\
P(\beta_{1})&=&\begin{cases}
              \text{uniform}\quad \text{if $-0.5<\alpha_{1}<0.5$ }\\
              $0$ \quad\text{otherwise}
          \end{cases}\nonumber\\
P(\beta_{2})&=& \begin{cases}
              \text{uniform}\quad \text{if $-2<\alpha_{2}<-0.5$ }\\
              $0$ \quad\text{otherwise}
              \end{cases}\nonumber\\
P(A_{V}^b)&=&\begin{cases}
              \text{uniform}\quad \text{if $0.5<A_{V}^b<1.1$ }\\
              $0$ \quad\text{otherwise}
              \end{cases}\nonumber\\ 
P(\sigma)&=&\begin{cases}
              \text{uniform}\quad \text{if $0<\sigma<1.5$ }\\
              $0$ \quad\text{otherwise}
              \end{cases}
\end{eqnarray}

(b) For $d$~--~$P/d$:
\begin{eqnarray}
P(\alpha)&=&\begin{cases}
              \text{uniform}\quad \text{if $0<\beta<2$ }\\
              $0$ \quad\text{otherwise}
              \end{cases}\nonumber\\
P(\beta_{1})&=&\begin{cases}
              \text{uniform}\quad \text{if $-1.4<\alpha_{1}<0$ }\\
              $0$ \quad\text{otherwise}
          \end{cases}\nonumber\\
P(\beta_{2})&=& \begin{cases}
              \text{uniform}\quad \text{if $-1<\alpha_{2}<0$ }\\
              $0$ \quad\text{otherwise}
              \end{cases}\nonumber\\
P(d^{b})&=&\begin{cases}
              \text{uniform}\quad \text{if $0.5<d^{b}<2.5$ }\\
              $0$ \quad\text{otherwise}
              \end{cases}\nonumber\\ 
P(\sigma)&=&\begin{cases}
              \text{uniform}\quad \text{if $0<\sigma<1$ }\\
              $0$ \quad\text{otherwise}
              \end{cases}
\end{eqnarray}

For the above two plots, Bayesian modelling is performed using a Python package called PyMC3 \citep{salvatieretal2016} via Markov Chain Monte-Carlo (MCMC) method using the Metropolis-Hastings sampling algorithm. The 95\%, 68\%, and 50\% confidence regions (CRs) predicted by the posterior distribution for each of the two plots are shown in Figure \ref{bayesianplots}.  From the figures, it is seen that the majority of the data fall well within 95\% CR predicted by the posterior. The 94\% highest  density intervals (HDI) of each parameter are plotted to represent the data lying within $2\sigma$ uncertainties. The posterior PDF of the parameters derived from the Bayesian modelling is given in Figure \ref{bayesianpdf}.  The parameters constrained through the  Bayesian analyses are reasonably more accurate than those obtained from the broken power-law alone. The constrained results from Bayesian analyses are listed in Table \ref{tab:bayesbrokenpowerlawparams}. 

To determine whether the broken power-law Bayesian model better represent the data over single-power-law Bayesian model,  we perform the leave one out ($loo$) cross-validation using $ArviZ$ module in {\sc Python}.  $loo$ cross validation estimates the expected log pointwise predictive density (elpd$_{loo}$) values for the broken-power law and single-power law bayesian models using the Pareto-smoothed importance sampling (PSIS) procedure. The standard error in elpd$_{loo}$ and the effective number of parameters are also estimated for both models.  $ArviZ$ then compares the broken power-law and single power-law models based on the standard errors in elpd$_{loo}$ values  and ranks the models from best to worst.  For both $A_{V}$~--~$P/A_{V}$ and $d$~--~$P/d$, $ArviZ$ ranked the broken power-law as a better model over the single power-law. 

\bsp
\label{lastpage}
\end{document}